\documentclass[amstex,useAMS,epsf,amssymb,usenatbib, twocolumn]{mn2e}
\usepackage{epsfig}
\usepackage{graphicx}
\usepackage{color}
\usepackage{amssymb,amsmath}
\usepackage{multicol}
\usepackage{soul}
\usepackage[normalem]{ulem}
\usepackage{url}
\usepackage{hyperref}
\usepackage{breakurl}
\usepackage{tabularx}

\newcolumntype{Y}{>{\centering\arraybackslash}X}

\newcommand{\gray}{$\gamma$-ray}
\newcommand{\grays}{$\gamma$ rays}

\title[Search for cyclical sources in the Fermi-LAT sky]{A search for cyclical sources of $\gamma$-ray emission on the period range from days to years in the Fermi-LAT sky}

\author[D. A. Prokhorov \& A. Moraghan]{D. A. Prokhorov$^{1}$\thanks{E-mail:phdmitry@gmail.com} and A. Moraghan$^{2}$\\
~\\
$^{1}$ School of Physics, Wits University, Private Bag 3, WITS-2050,
Johannesburg, South Africa\\
$^{2}$ Academia Sinica Institute of Astronomy and Astrophysics, P.O.
Box 23-141, Taipei 106, Taiwan }

\date{Accepted .....
      Received ..... ;
      in original form .....}

\pagerange{\pageref{firstpage}--\pageref{lastpage}} \pubyear{2017}

\begin{document}

\maketitle

\begin{abstract}
A systematic search for cyclical sources of $\gamma$-ray emission on
the period range from days to years in the Fermi-LAT sky is
performed. Looking for cyclical emission, the sky is binned into
equal-area pixels and the generalised Lomb-Scargle periodogram is
computed for each of these pixels. The search on the period range
between 2.5 and 30 days in the Galactic plane confirms periodicities
of three binaries, LSI +61$^{\circ}$303, LS 5039, and 1FGL
J1018.6-5856. The all-sky search on the period range between 30 days
and 2.5 years confirms periodicities of three blazars, PG 1553+113,
PKS 2155-304, and BL Lacertae. Evidence for periodic behaviours of
four blazars, 4C +01.28, S5 0716+71, PKS 0805-07, and PKS 2052-47,
are presented. Three of these blazars, 4C +01.28, PKS 0805-07, and
PKS 2052-47, are located at high redshifts. These three sources are
potential candidates to binary systems of supermassive black holes
provided that major galaxy mergers are more frequent and that
galaxies are more gas-rich at high redshifts.
\end{abstract}


\section{Introduction}

Two classes of sources are known to be cyclical emitters in the HE
(0.1-100 GeV) \gray{} regime on the period range from days to years.
The first class consists of binaries showing orbitally modulated
\gray{} emission, including LS 5039 \citep[3.9 day
period,][]{bnrLS5039}, 1FGL J1018.6-5856 \citep[16.6
days,][]{bnrJ1018}, LS I +61$^{\circ}$303 \citep[26.5
days,][]{bnrLSI61} in the Milky Way, and CXOU J053600.0-673507
\citep[10.3 days,][]{bnrLMC} in the Large Magellanic Cloud. The
second class consists of blazars, showing strong evidence for
quasi-periodic modulations in \grays{}, including PG 1553+113
\citep[2.2 year period,][]{pg1553}, PKS 2155-304 \citep[1.7
years,][]{pks2155p1, pks2155p2}, and BL Lacertae \citep[1.9
years,][]{bllac}. Modulated \gray{} signals from these
sources were detected by {\it{Fermi}}-LAT. This pair-conversion
\gray{} space-based telescope \citep[][]{atwood2009} normally
operates in sky-survey mode where the whole sky is observed every 3
hr. It provides a regular and uniform view of \gray{} sources
allowing us to monitor binaries and blazars on a daily basis.
Multi-wavelength observations provide independent confirmations of
periodic signals. Thus, the \gray{} binaries also emit signals with
the same periods at other frequencies \citep[e.g., periodic TeV
emission with a 3.9 day period from LS 5039 detected by H.E.S.S.,
see][]{aharonian2006}, while the blazar, PG 1553+133, emits an
optical signal of period consistent with that detected in HE
\grays{} \citep[][]{pg1553}.

Hundreds of interacting binaries are X-ray emitters
\citep[e.g.,][]{Liu2006, Liu2007}, but very few binaries produce
detectable \gray{} emission. Detected \gray{} binaries are systems
composed of a massive ($>$10 solar masses) star and a compact
object. The latter is either a neutron star or a black hole. The
periods determined from HE \gray{} data are compatible with binary
periods obtained from radial velocity measurements
\citep[e.g.][]{aragona2009}. Physical mechanisms responsible for
periodic HE \gray{} radiation likely invoke inverse Compton (IC)
scattering from HE electrons or electron-positron pairs \citep[for a review,
see][]{dubus}. Target photons for IC processes are provided by the
massive star and their radiation density varies along the eccentric
orbit. HE electrons accelerated in the vicinity of the compact
object up-scatter the stellar radiation. Since the angle at which an
observer sees the star and HE electrons changes with the orbit,
orbital modulation is expected.

\gray{} blazars are the most numerous type of \gray{} sources
\citep[][the \textit{Fermi}-LAT 3FGL catalog includes $>$1500
blazars]{3fgl}. However, most of the \gray{} blazars show erratic
variability on a wide range of timescales \citep[][]{2fav} without
any evidence for periodicity. Blazars are active galactic nuclei
(AGN) emitting a relativistic jet that is pointing very nearly along
Earth's line of sight. At HE energies, blazar emission is often
satisfactorily modelled as IC scattering of seed photons by HE
electrons. Seed photons can be synchrotron photons produced in the
jet itself \citep[][]{maraschi92, bloom96} or photons from an
external source, such as the accretion disk, broad-line region, or
dusty infrared torus \citep[][]{sikora1994}. A temporal analysis
plays an important role in studying the origin of \gray{} emission
from blazars. In some cases, the separation of \gray{} events
recorded during AGN flares from those during quiescent states can be
used as a method of disentangling contributions from various
emission components \citep[e.g.,][]{paper1}. Various types of time
variabilities in blazars are conceivable. Periodic \gray{} emission
from blazars is expected if supermassive binary black holes are
present at their centres \citep[e.g.,][]{rieger2007}. The
geometrical origin of periodicity in \gray{} blazars, involving jet
precession, intrinsic jet rotation, or helical structures in jets,
is also possible \citep[e.g.,][]{rieger2004, mohan2015, raiteri2015,
cavaliere2017, sobacchi2017}. Quasi-periodic fluctuations in the
energy outflow efficiency in the magnetically arrested accretion
regime \citep[][]{tchekhovskoy2011} suggest another potential
mechanism for cyclic variability.

A supermassive black hole (SMBH) is believed to reside at the centre
of every large galaxy \citep[e.g.,][]{Kormendy1995}. In hierarchical
$\Lambda$CDM cosmology, galaxies grow and evolve through mergers and
accretion of smaller substructures \citep[e.g.,][]{White1991}. If
more than one of the merging galaxies contain an SMBH, two or more
SMBHs will be present in their resultant galaxy
\citep[e.g.,][]{Kulkarni2012}. Gravitationally bound binary SMBH
systems are expected to be a product of mergers of galaxies and form
when the separation between the SMBHs gradually shrinks
\citep[][]{Begelman1980}. A binary SMBH system can efficiently be
formed in gas-dominated galaxies \citep[e.g.,][]{Callegari2009}.
Since major mergers of galaxies are more frequent at high redshifts
\citep[][]{Volonteri2009} when galaxies are also more gas-rich
\citep[][]{Tacconi2010}, the number of SMBH binaries increases
rapidly with increasing redshift. SMBH binaries themselves are
expected to be intrisically rare and a detection of quasi-periodic
emission from a high-redshift blazar possibly modulated via the
dynamics of gravitationally bound SMBHs can provide evidence for the
existence of a population of sub-pc separation SMBH binaries
\citep[e.g.,][]{Bogdanovic2015}. The previously claimed
quasi-periodic $\gamma$-ray signals (that we also confirm) are from
blazars, PG 1553+133 (z=0.360), PKS 2155-304 (z=0.116), and BL
Lacertae (z=0.068), at low redshifts\footnote{the redshifts of
blazars are taken from the NASA/IPAC Extragalactic Database}. In
this paper, we pay our attention to cyclic $\gamma$-ray blazars
during the early epoch of $0.7\lesssim z\lesssim2.0$.

The all-sky monitoring capability of {\it{Fermi}}-LAT is crucial to
perform a search for new cyclical \gray{} sources. Using the
\textit{Fermi}-LAT data, we perform a systematic search for cyclical
\gray{} binaries with periods between 2.5 days and 30 days in the
Galactic plane, $-10^{\circ}<b<+10^{\circ}$, and a systematic
all-sky search for cyclical \gray{} blazars with periods between 30
days and 2.5 years. To perform these investigations, the \gray{} sky
is binned into equal-area pixels using the HEALPix package
\citep[][]{Gorsky2005} and the generalised Lomb-Scargle (GLS)
periodogram \citep[][]{Zechmeister2009} is applied to establish
periodic sources. The results of the search confirm the
periodicities of \gray{} binaries LS 5039, 1FGL J1018.6-5856, and LS
I +61$^{\circ}$ 303, and \gray{} blazars PG 1553+113, PKS 2155-304,
and BL Lacertae. In addition, new candidates in quasi-periodic
\gray{} sources amongst blazars are found and include 4C +01.28, PKS
0805-07, PKS 2052-47, and S5 0716+71. Three of these $\gamma$-ray
sources, 4C +01.28, PKS 0805-07, and PKS 2052-47, are located at
high redshifts (z=0.890, 1.837, and 1.489, respectively) and
significantly more distant than PG 1553+113, PKS 2155-304, and BL
Lacertae. Therefore, the blazars, 4C +01.28, PKS 0805-07, and PKS
2052-47, are potential candidates to binary SMBH systems. Our
systematic analysis is also a powerful tool to separate
astrophysical periodic \gray{} signals from the known instrumental
effects, in addition to effects caused by the Sun and the Moon
moving through the sky, that all lead to an extrinsic \gray{}
modulation.

\section{\textit{Fermi}-LAT observations and time series analysis}

The Fermi satellite was launched on 2008 June 11 into a
nearly circular Earth orbit with an altitude of 565 km
and an inclination of 25.6$^{\circ}$.
The orbital period is 96.5 minutes, and the orbit has a precession period of 53.4 days
(so the RA and Dec of the orbit poles trace a 25.6 degree circle on the sky
every 53.4 days).
The Fermi satellite has two instruments onboard, the Large Area
Telescope (LAT) pair-conversion detector and the Gamma Ray Burst Monitor
(GBM).
The \textit{Fermi}-LAT instrument \citep[][]{atwood2009} provides coverage over
the energy range from $\sim$20 MeV to several hundreds of GeV.
It has a large field of view ($\approx$20\% of the sky), and
has been scanning the entire sky since it began routine science operation on
2008 August 4.
It provides an angular resolution per single event of
5$^{\circ}$ at 100 MeV, narrowing to 2$^{\circ}$ at 300 GeV,
and further narrowing to 0.15$^{\circ}$ at 10 GeV \citep[][]{atwood2013}.

The Fermi-LAT Pass 8 data is downloaded from the Fermi Science
Support
Center\footnote{\burl{http://heasarc.gsfc.nasa.gov/FTP/fermi/data/lat/weekly/photon/}}.
The data acquired during each week of the Fermi's science mission
are contained in weekly data files. Pass 8 is an event-level
reconstruction analysis framework \citep[][]{atwood2013} applied to
the data taken by the \textit{Fermi}-LAT, which results in an
increase in \gray{} acceptance by 20-40\% with respect to the
previous Pass 7 reprocessed data release. For the data analysis, the
\texttt{Fermi Science Tools}~v10r0p5
package\footnote{\burl{https://fermi.gsfc.nasa.gov/ssc/data/analysis/software/}}
and \texttt{P8R2\_SOURCE\_V6} instrument response functions are
used. \texttt{Pass 8 SOURCE class} photon data (evclass=128)
spanning 7.8 years (MET~239557417 - 485913604) with energies between
300 MeV and 500 GeV are selected. The \texttt{SOURCE} event class is
tuned to balance statistics with background flux for long-duration
point source analysis. By excluding events with reconstructed
energies below 300 MeV, we tighten the point spread function (PSF).
Contamination from the \gray{}-bright Earth's limb is avoided by
removing all events with zenith angle $>$90$^{\circ}$. The
recommended quality cuts (\texttt{DATA QUAL==1 \&\& LAT CONFIG==1})
are applied. The data is binned into time intervals of 12 hours and
is binned using the \texttt{HEALPix} package into a map of
resolution $N_{\mathrm{side}}$=32 in Galactic coordinates with
`RING' pixel ordering. With these settings, the total number of
pixels is equal to 12288 and the area of each pixel is 3.4 sq. deg.
The resolution of the map is chosen according to the size of the
\textit{Fermi}-LAT PSF above 300 MeV. To compute the exposure for
each pixel, the standard tools \texttt{gtltcube} and
\texttt{gtexpcube2} are used. To correct the livetime for the zenith
angle cut, the ``zmax'' option on the command line is used. For each
pixel and each time interval, the number of photons are counted, the
corresponding value of exposure is computed, and the integral
\gray{} flux is calculated. \textit{Fermi}-LAT light curves are
created through aperture photometry to provide a model independent
measure of the flux. The time bins used in this paper are
sufficiently long (of 12 hours at least) and, therefore, barycenter
corrections of the arrival times of the \gray{} photons are
neglected. The search for cyclical $\gamma$-ray emission in binned
equal-area portions of the sky has the advantage that it minimises
the number of assumptions required for quasi-periodicity searches
(e.g., it does not rely on both the distribution of $\gamma$-ray
sources in the sky and the foreground diffuse model). However, the
minimisation of the assumptions in turn leads to a decrease in the
sensitivity of the method due to the possible leakage of $\gamma$
rays from a source to neighbouring pixels and due to the
contamination of a $\gamma$-ray signal from the strongest
$\gamma$-ray source in the pixel by other $\gamma$-ray sources
belonging to the same pixel.

The Lomb-Scargle periodogram is a common statistical tool used in
time series analysis to search for periodicities
\citep[][]{Lomb1976, Scargle1982}. The problem in question is to
establish the existence of a periodic signal, despite the presence
of noise, where noise is random variations in the source flux. The
method is equivalent to fitting sinusoidal waves,
$y=\alpha\mathrm{cos}(\omega t)+\beta\mathrm{cos}(\omega t)$, in
searches for periodic \gray{} signals. 
In this paper, the GLS periodogram \citep[][]{Zechmeister2009} is
used. Compared with the Lomb-Scargle periodogram, the GLS
periodogram takes the measurement errors into account. It also
introduces an offset, $c$, resulting in a generalisation of the
periodogram to the equivalent of fitting a sinusoidal wave plus a
constant, i.e., $y=\alpha\mathrm{cos}(\omega
t)+\beta\mathrm{cos}(\omega t)+c$. In the context of Galactic
\gray{} binaries, the diffuse Galactic \gray{} emission produced
through pion decay \citep[][]{Porter2012} contributes to a constant
term, whereas in the context of \gray{} blazars, it is the isotropic
diffuse \gray{} background \citep[][]{xgal2015} that contributes to
a constant term. The known \gray{} binaries have periods between 3.9
days and 26.5 days and the known cyclic \gray{} blazars have periods
between 1.7 years and 2.2 years, and therefore enclose many 12-hour
time intervals. Time bins of a 12-hour and 1-day duration are used
to search for \gray{} binaries in the Galactic plane,
$-10^{\circ}<b<+10^{\circ}$. Longer time bins of a 3-day, 7-day, and
14-day duration are used for the all-sky search for cyclic \gray{}
blazars. These time bins contain a whole number of 12-hour
intervals. Typically, the average number of \gray{} events per time
bin is $>10$. In this paper, the python module,
\texttt{astroML}\footnote{\burl{http://www.astroml.org/}}
\citep[][]{astroML2012}, is used for a time series analysis. The GLS
periodograms are computed and the positions of peaks in the GLS
periodograms are obtained using the tool,
\texttt{astroML.time\_series.lomb\_scargle}. The chance probability
of finding a peak in the GLS periodogram higher than the observed
value is also estimated using this tool and is cross-checked using
an alternative method,
\texttt{astroML.time\_series.lomb\_scargle\_bootstrap}. Both these
estimates of the chance probability are computed under a condition
of the exclusive presence of white (i.e., flat-spectrum) noise.
$\gamma$-ray light curves of blazars are generally characterised by
power-law noise with the power spectrum density (PSD) decreasing
with frequency \citep[e.g.,][]{ferminoise1, ferminoise2,
ferminoise3}. In this case, the chance probability is lower than
that for the white noise distribution because of more power in
longer time scales. Therefore, firstly we compute the chance
probability under a condition of white noise for each pixel in the
sky; secondly comparing the computed chance probability with the
probability threshold we select pixels with possible $\gamma$-ray
cyclic candidate sources; and thirdly if the candidate source is
associated with a blazar, we calculate the power spectrum density of
the source and perform simulations to check reliability of the
source quasi-periodicity.

\section{Results of the search for cyclical \gray{} sources}

In this Section, the results of the search for \gray{} sources in
the \textit{Fermi}-LAT sky are presented. To establish a periodicity
of a \gray{} source, the following requirements must be satisfied.
The highest observed peak in the GLS periodogram must have a very
small probability to be found by chance under a condition of white
noise. In other words, the chance probability of
$<5.7\times10^{-7}$, corresponding to a $>$5$\sigma$ confidence
level, must be satisfied for detection of periodic-like behaviour.
Furthermore, to ensure that the conclusion on a source periodicity
does not depend on the bin width of the light curve, the periodicity
detection found using a one time binning scheme must be confirmed
using an alternative time binning scheme, and the periods found
using these two different time binning schemes must be consistent.
To be associated with a \gray{} source, a pixel showing a
periodic-like behaviour must contain a source from the
\textit{Fermi}-LAT 4-year point source (3FGL) catalog. The
associated \gray{} sources are additionally analysed by selecting
\grays{} within a $1^{\circ}$ radius (and also within a more
extended region of a $2^\circ$ radius) from the 3FGL source position
and at energies above 300 MeV. The additional analyses by means of
the GLS periodogram must also show the presence of a periodicity
above a 5$\sigma$ level (under a condition of white noise) in order
to establish a periodicity of the signal from the 3FGL source. Using
150000 bootstrap resamplings, the presence of a periodicity of the
associated \gray{} source at a $>4\sigma$ confidence level is
cross-checked. If the associated $\gamma$-ray source is a blazar,
then we compute its PSD by means of the
\texttt{astroML.fourier.PSD\_continuous} tool, fit the computed PSD
with a power-law function to calculate the power-law index, and use
simulations to assess the chance probability to observe the
corresponding peak in its GLS periodogram under a condition of
power-law noise by means of the
\texttt{astroML.time\_series.generate\_power\_law} tool. Finally, a
check if a periodic-like behavior of the source can be caused by the
known instrumental effects or periodic \gray{} background is
performed. The \gray{} sources satisfying all these requirements are
shown in Table \ref{Tab}. The 1st column of Table \ref{Tab} shows
the pixel number according to the `RING' pixel ordering, the 2nd
column shows the signal period in units of days, the 3rd and 4th
columns show the name of the associated sources and its name in the
3FGL catalog. The 5th column shows the source type. The 6th shows
the chance probability that the detected periodic signal belongs to
the white (flat) noise. The 7th column shows the best-fit value of
the PSD slope. The 8th shows the chance probability that the
detected periodic signal belongs to the noise with the given
power-law PSD. The results of searches for \gray{} binaries and
blazars are described in the next two paragraphs. A short discussion
about the instrumental effects is given at the end.

\begin{table*}
\centering \caption{Cyclical sources of HE \gray{} emission. The values shown in bold type
are obtained from a more sensitive analysis of the 3FGL source belonging to the pixel.}
\begin{tabular}{ | c | c | c | c | c | c | c | c | c | c |}
\hline
Pixel & Period & Source name & 3FGL name & Type & Probability & PSD slope & Probability \\
number & (days) &  &  &  & (white noise) &  & (power-law noise) \\
\hline
1269 & 445 & 4C +01.28 & J0158.5+0133 & blazar & $3\times10^{-7}$\% & 0.64 & \textbf{0.9\%}\\
1867 & 798 & PG 1553+113 & J1555.7+1113 & blazar & $<1\times10^{-10}$\% & 0.53 & $<$0.1\%\\
3187 & 346 & S5 0716+71 & J0721.9+7120 & blazar & $4\times10^{-10}$\% & 0.57 & 0.1\%\\
4753 & 658 & PKS 0805-07 & J0808.2-0751 & blazar & $2\times10^{-7}$\% & 0.77 & 6.7\%\\
6000 & 27.0 & LSI +61$^{\circ}$303 & J0240.5+6113 & binary & \textbf{$<1\times10^{-10}$\%} & - & - \\
6214 & 3.9 & LS 5039 & J1826.2-1450 & binary & $<1\times10^{-10}$\% & - & - \\
6309 & 16.5 & 1FGL J1018.6-5856 & J1018.9-5856 & binary & \textbf{$7\times10^{-7}$\%} & - & - \\
7265 & 698 & BL Lacertae & J2202.7+4217 & blazar & $1\times10^{-10}$\% & 0.74 & 0.9\%\\
10173 & 637 & PKS 2052-47 & J2056.2-4714 & blazar & $3\times10^{-10}$\% & 0.69 & \textbf{0.4\%}\\
10992 & 644 & PKS 2155-304 & J2158.8-3013 & blazar & $<1\times10^{-10}$\% & 0.67 & 0.2\%\\
\hline
\end{tabular}
\label{Tab}
\end{table*}

\begin{figure*}
\centering
  \begin{tabular}{@{}cc@{}}
    \includegraphics[angle=0, width=.48\textwidth]{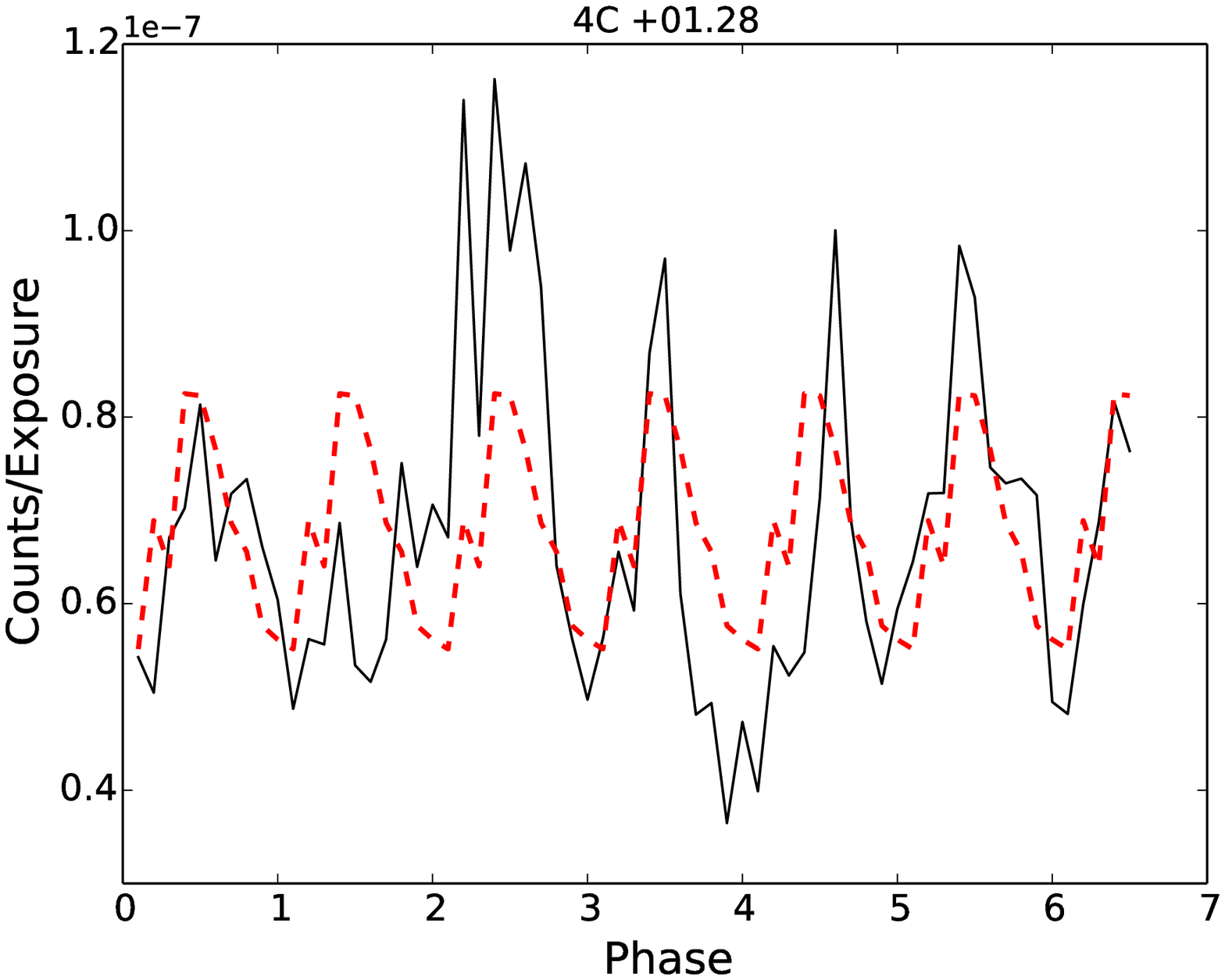}
    & \ \ \ \ \ \ \
    \includegraphics[angle=0, width=.48\textwidth]{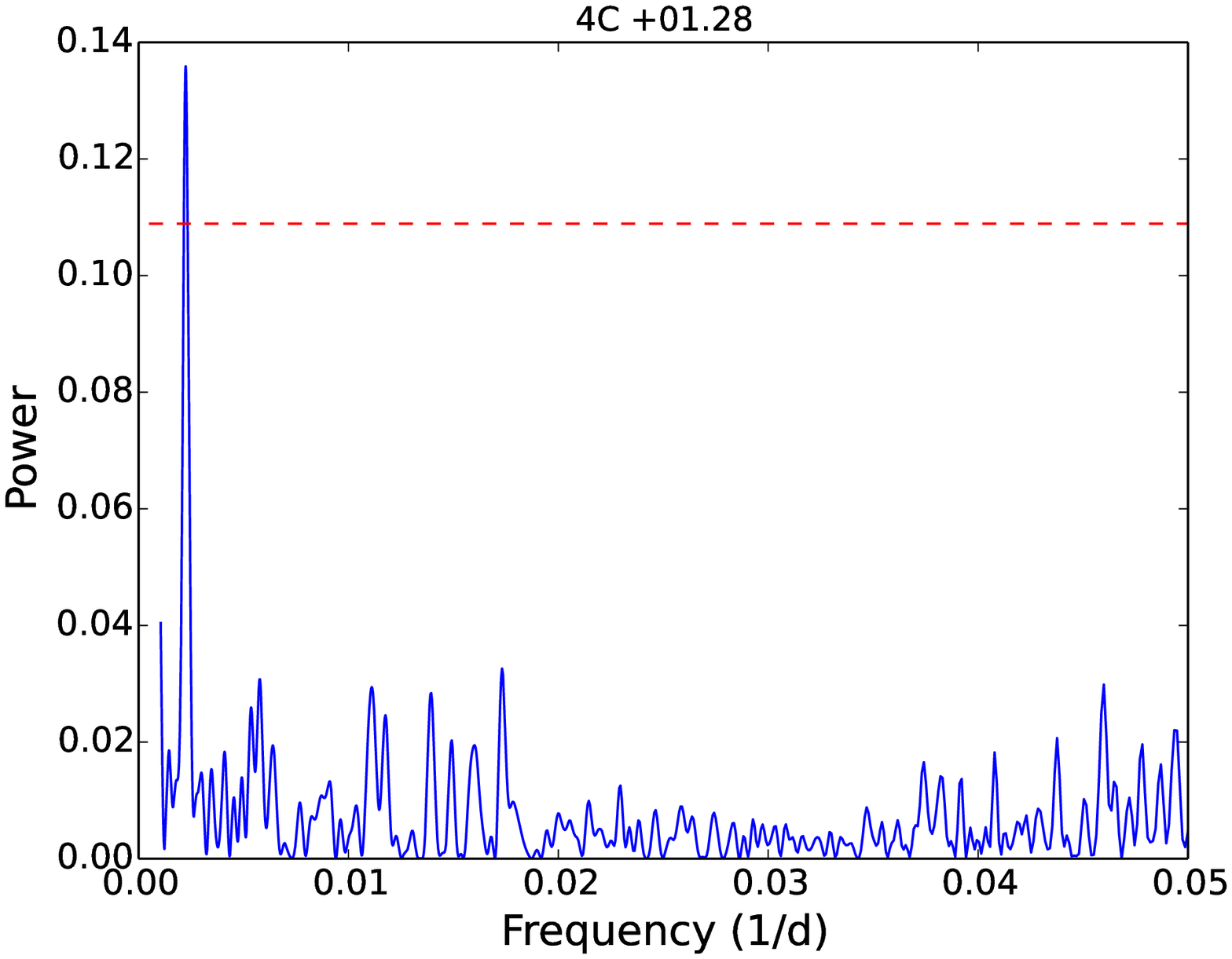}\\
    \includegraphics[angle=0, width=.48\textwidth]{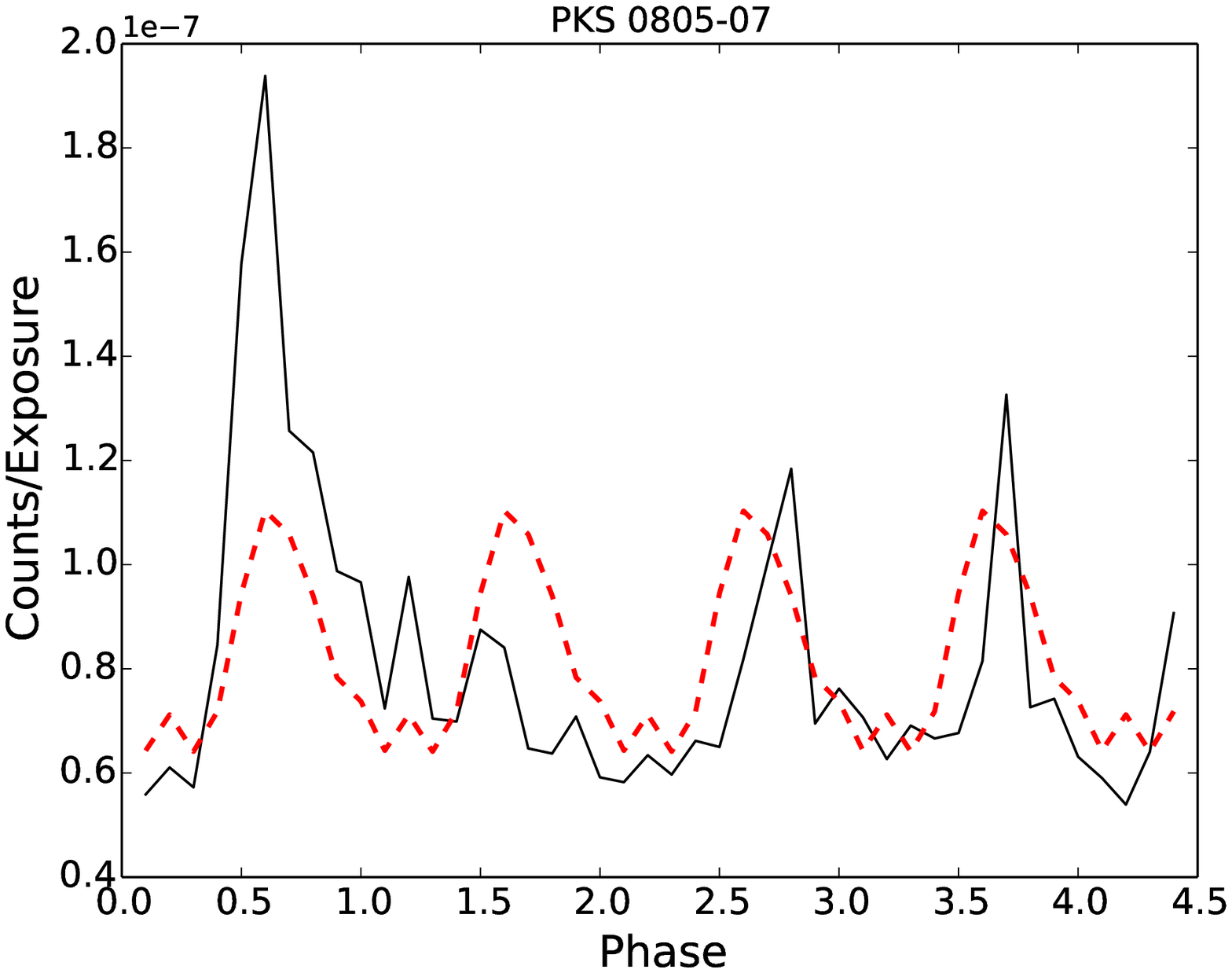}
    & \ \ \ \ \ \ \
    \includegraphics[angle=0,
    width=.48\textwidth]{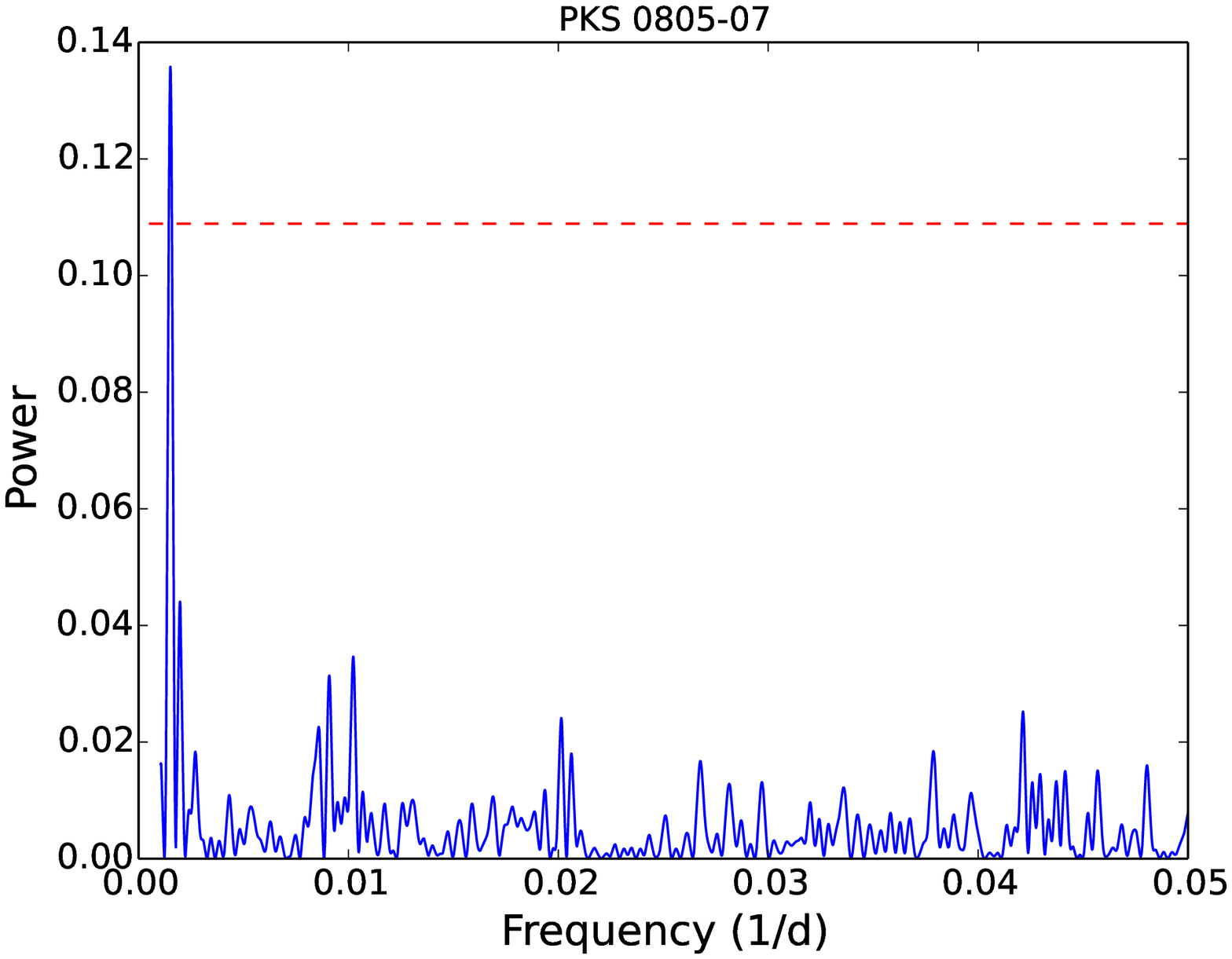}\\
    \includegraphics[angle=0, width=.48\textwidth]{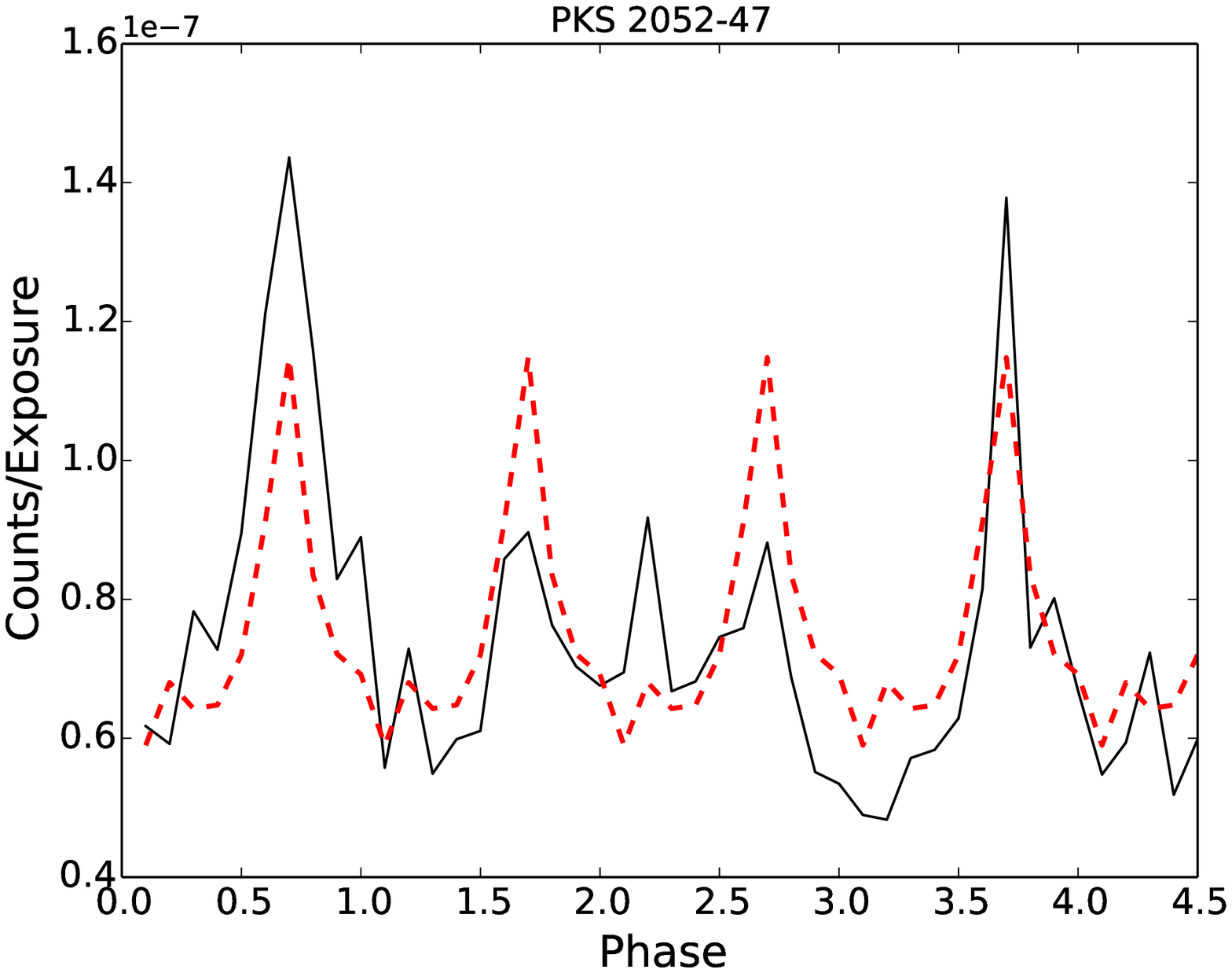}
    & \ \ \ \ \ \ \
    \includegraphics[angle=0,
    width=.48\textwidth]{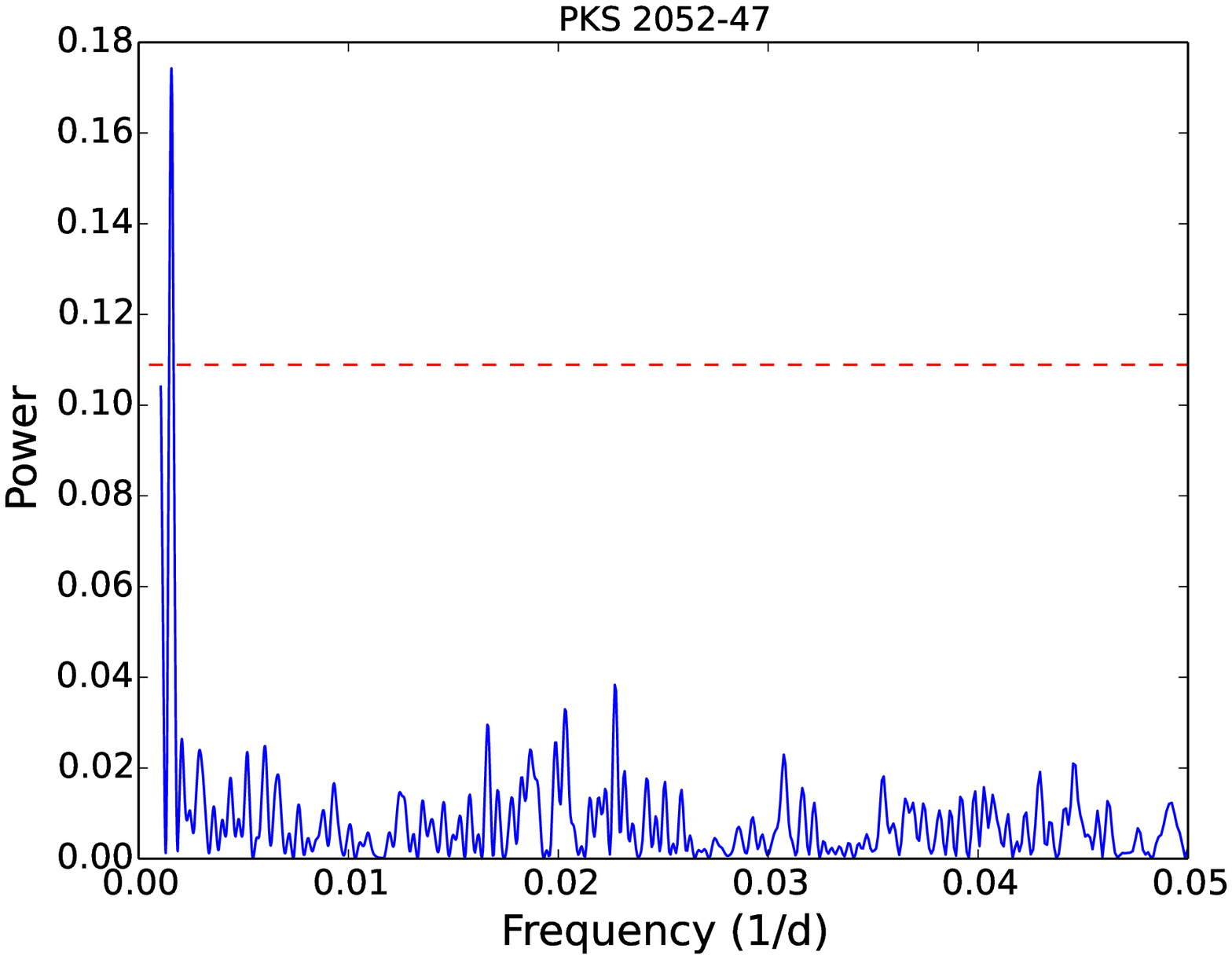}
   \end{tabular}
  \caption{The left panel shows the periodic-like variations of the measured 
flux of the three high-redshift candidates to quasi-periodic $\gamma$-ray
sources as a function of the phase (solid line) and the average flux
within each phase (dashed line). The right panel shows the GLS
periodogram for these sources, where the dashed line shows the level
corresponding to the chance probability of $<5.7\times10^{-7}$ under
a condition of white noise.}  \label{F1}
\end{figure*}

The analysis performed on the period range between 2.5 and 30 days
shows that the pixels with numbers 6000 and 6214 satisfy all the
above requirements. These two pixels contain \gray{} sources, LSI
+61$^{\circ}$303 and LS 5039. The $\gamma$-ray flux above 300 MeV
(taken from the 3FGL catalog) from LSI +61$^{\circ}$303 is fourteen
times higher than that of the other 3FGL source in the former pixel,
while the flux above 300 MeV from LS 5039 is eleven times higher
than the sum of fluxes of the two other sources belonging to the
latter pixel. Therefore, we identify LSI +61$^{\circ}$303 and LS
5039 with $\gamma$-ray cyclic emission in these two pixels. The
computed periods, 27.0 and 3.9 days, are consistent with those
reported by \citet[][]{bnrLSI61, bnrLS5039}. Only these two pixels
in the Galactic plane show a periodic behaviour at a $>5\sigma$
confidence level on this period range in two different time binning
schemes (12-hour and 1-day time bins). The periodic behaviour of
these two binaries was previously established in \grays{} during the
1st year of the \textit{Fermi}-LAT mission. The pixel with number
6309 is the only other pixel in the Galactic plane showing a
periodic behaviour at a $>4\sigma$ confidence level in different
time binning schemes. This pixel contains the known \gray{} binary,
1FGL J1018.6-5856. The highest peak for the pixel number 6309,
corresponds to a period of 16.5 days which is consistent with that
found for binary, 1FGL J1018.6-5856, by \citet[][]{bnrJ1018}. The
additional analysis of \grays{} within a $1^{\circ}$ radius from the
position of the
 source 3FGL J1018.9-5856 contained in this pixel (and corresponding
to 1FGL J1018.6-5856), at energies above 300 MeV, results in a
larger periodogram peak, shows that the chance probability that the
detected periodic signal belongs to the noise decreases from 0.001\%
to $7.0\times10^{-7}$\%, and therefore shows a periodic behaviour at
a $>5\sigma$ confidence level. This binary is also included in Table
\ref{Tab}. The pixel containing this binary also contains six other
3FGL sources and the ratio of the flux above 300 MeV from 3FGL
J1018.9-5856 to the sum of the other sources' fluxes is about 0.8.
Therefore, centering of the aperture at the binary position
increases the significance of a periodic behaviour. Similarly, the
additional analysis of $\gamma$-ray emission from LSI
+61$^{\circ}$303 shows that the chance probability that the detected
periodic signal belongs to the noise decreases from
$4.0\times10^{-7}$\% to $<1.0\times10^{-10}$\% when the aperture
with a 1 deg radius centred on the source is used. The performed
search confirms the previous results about periodicities of three
\gray{} binaries belonging to the Galactic plane. No new candidates
to periodic \gray{} sources with properties similar to these
binaries is established in the Galactic plane.

\begin{figure*}
    \centering
    \includegraphics[width=0.7\textwidth]{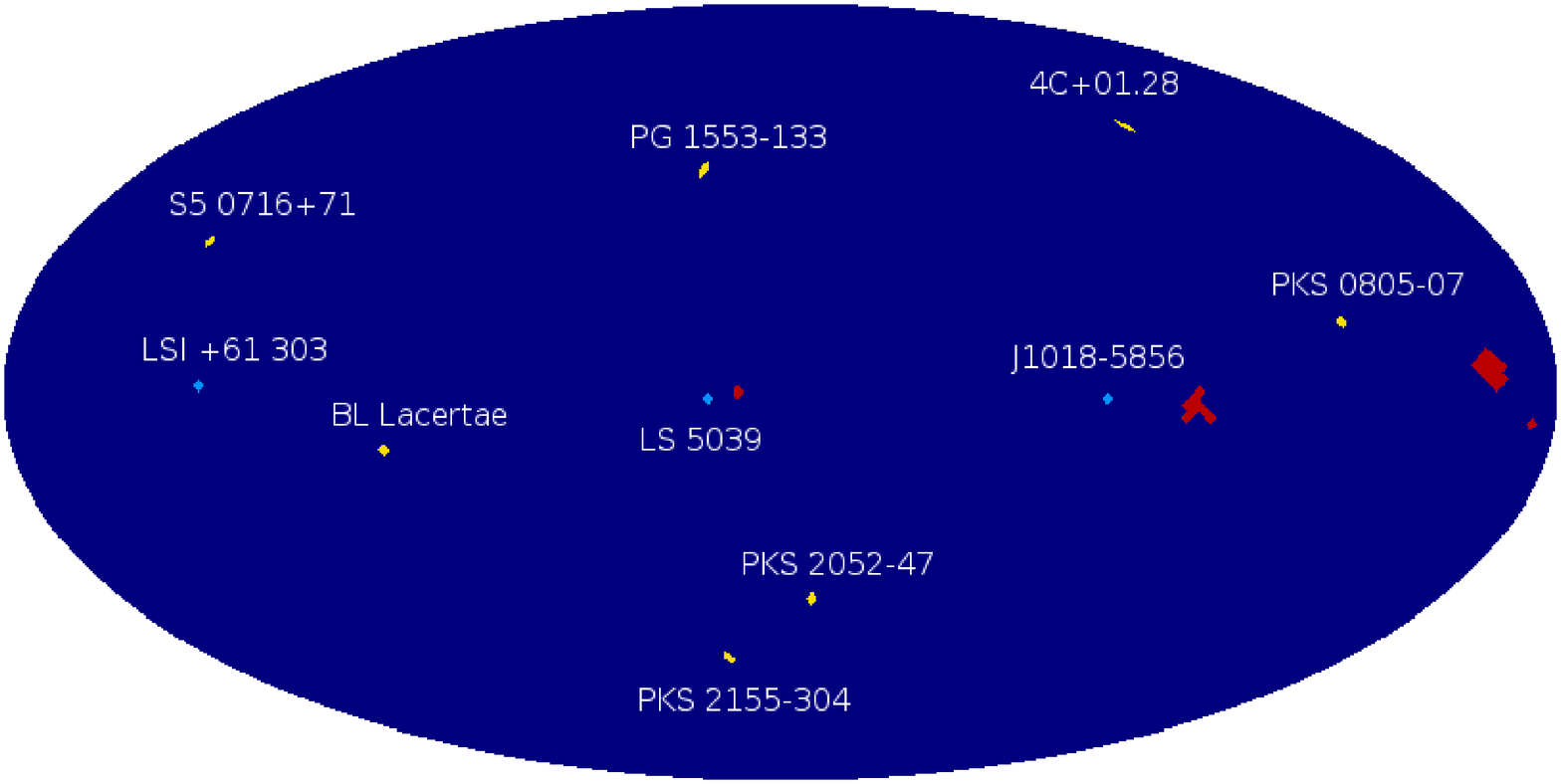}
    \caption{Figure shows periodic-like \gray{} sources in the \textit{Fermi}-LAT
    sky. These include three binaries (the corresponding pixels are shown in blue)
    and seven blazars (shown in yellow). Modulated \gray{} signals from the pixels
    shown in red are due to instrumental effects, see the text for details.
    This figure can be seen in colour in the online version.}
    \label{Fig2}
\end{figure*}

The analysis performed on the period range between 30 days and 2.5
years shows that seven pixels with numbers 1269, 1867, 3187, 4753,
7265, 10173, and 10992 satisfy all the periodicity requirements.
These seven pixels contain seven \gray{} blazars. For five of these
pixels there is a one-to-one correspondence between pixels and 3FGL
sources, while the remaining two pixels, 1269 and 4753, contain two
3FGL sources each. One of the 3FGL sources in each of these pixels
is a strongly dominant contributor to the total flux from the pixel
above 300 MeV (with a fraction of 90\% and 97\%, respectively). It
allows us to avoid ambiguity in quasi-periodic source
identification. Three of these blazars are PG 1553+113, BL Lacertae,
and PKS 2155-304, whose \gray{} periodicities were previously
reported in the literature. The computed periods (see Table
\ref{Tab}) agree with those previously reported. These three blazars
show the highest statistical chances to be periodic \gray{} sources.
The four other blazars, 4C +01.28, S5 0716+71, PKS 0805-07, and PKS
2052-47, are new candidates to quasi-periodic \gray{} blazars. Three
of the four candidates are located at high redshifts of z=0.89 (4C
+01.28), 1.84 (PKS 0805-07), and 1.49 (PKS 2052-47), and are much
more distant than the previously reported quasi-periodic \gray{}
blazars. Taking that binary SMBH systems are expected to be more
frequent at high redshifts into account, these distant blazars are
potential candidates to binary systems of SMBHs. The fourth
candidate is S5 0716+71 and is located at redshift of 0.31. The LS
periodogram analysis of the optical data for this blazar reveals two
significant peaks at 359 and 63 days \citep[][]{Rani2013}. The
former peak is at the frequency close to the frequency of the peak
in the GLS periodogram of the $\gamma$-ray data. But it is hard to
provide strong, independent, supporting evidence in favour of a
quasi-periodic behaviour at this frequency due to limited number of
observed cycles in the optical data \citep[][]{Rani2013}. To
illustrate possible quasi-periodic behaviours of the three new
high-redshift candidates, the light curves are binned into time
intervals of width one tenth of the corresponding source period.
Figure \ref{F1} shows cyclical variations of the measured flux
(i.e., counts/exposure) as a function of the phase for each of these
\gray{} blazars. The average flux within each phase derived from the
\textit{Fermi}-LAT observations is overplotted as a dashed line.
This figure also shows the GLS periodograms for these sources.
Although, cyclic behaviors of these four candidates are less
statistically significant than those of PG 1553+113, BL Lacertae,
and PKS 2155-304 (see Table \ref{Tab}), multi-wavelength analyses of
these quasi-periodic source candidates are of interest to check if
periodic-like signals are present at other frequencies. Taking the
periods of the seven blazars showing the possible quasi-periodic
behaviours into account, one can expect to observe their next
several \gray{} emission cycles in the coming five years. The
locations of these ten sources (three binaries + seven blazars),
showing periodic-like $\gamma$-ray emission in the sky are
visualised in Fig. \ref{Fig2}.

To assess the chance probability to observe the corresponding peaks
in the GLS periodograms of the seven blazars under a condition of
power-law noise, we calculated the PSD for these sources for the
\textit{Fermi}-LAT weekly light curves. Using the
\texttt{optimize.leastsq} tool from the \texttt{SciPy} library, we
applied least-square fits to the PSD assuming a power-law
$\mathrm{PSD}\propto\nu^{-\alpha}$, where $\nu$ is the frequency.
The best-fit values of the PSD slopes are shown in Table \ref{Tab}.
For PKS 2155-304 and BL Lacertae, the best-fit PSD slope values of
$0.67$ and $0.74$, respectively, are compatible with $0.58\pm0.33$
and $0.41\pm0.47$, reported by
\citet[][]{ferminoise2}, and with $0.64\substack{+0.79 \\
-0.50}$ and $0.93\substack{+0.18 \\ -0.14}$, reported by
\citet[][]{ferminoise3}. We use the
\texttt{astroML.time\_series.generate\_power\_law} tool based on the
algorithm from \citet[][]{Timmer1995} to simulate power-law noise
light curves with the given PSD slope. For each of the simulated
light curves, we compute the GLS periodogram. We calculate the
chance probabilities to observe the peaks in the GLS periodograms of
the seven blazars by comparing their power peak values with those
obtained in these simulations. The calculated chance probabilities
that the detected periodic signal belongs to the power-law noise are
shown in Table \ref{Tab}. This analysis shows that the power peaks
in the periodograms of PG 1553+133, PKS 2155-304, and BL Lacertae
have $<1$\% chance to be caused by a statistical fluctuation. For
the three high-redshift candidates to quasi-periodic $\gamma$-ray
sources, the power peaks of 4C +01.28 and PKS 2052-47 have $2.7$\%
and $1.0$\% chance to be caused by a statistical fluctuation, while
the chance of the power peak of PKS 0805-07 is $6.7$\%. Though the
chance of the power peak of PKS 0805-07 is above 5\%, this
$\gamma$-ray blazar is the most distance source (at z=1.837) amongst
the four new candidates. The accumulation of more $\gamma$-ray data
is needed to re-access its quasi-periodicity. We checked and found
that the results obtained from equal-area portions binned with
\texttt{HEALPix} and containing these sources, and from circular
regions of $1^{\circ}$ (or $2^{\circ}$) radius around these sources
are consistent.

To reduce uncertainty caused by both the diffuse background and
background due to the presence of $\gamma$-ray sources projected
near the new candidates to cyclic $\gamma$-ray blazars, we performed
a likelihood analysis of these four blazars using the standard
\texttt{Fermi Science Tools} package. We selected events from 300
MeV to 500 GeV in a circular region of interest of $20^{\circ}$
radius centred on the position of each of the blazars. The data was
binned in thirty equal logarithmically spaced energy intervals. To
model the Galactic and isotropic background diffuse emission, we
used the templates, \texttt{gll\_iem\_v06} and
\texttt{iso\_P8R2\_SOURCE\_V6\_v06}. The other cuts applied to the
\textit{Fermi}-LAT data are identical to those used in Sect. 2. We
included $\gamma$-ray sources from the 3FGL catalog to model the
data within the regions of interest. A binned maximum likelihood
model fit method was applied to each weekly time bin to derive the
light curves of the integrated flux (E$>$300 MeV). We computed the
GLS periodograms of these sources and found that their power peaks
are at 442 days for 4C +01.28, at 340 days for S5 0716+71, at 676
days for PKS 0805-07, and at 642 days for PKS 2052-47. These
positions of the GLS power peaks are compatible with those obtained
above (see Table \ref{Tab}). Three of these four power peaks are
significant at $>5\sigma$ confidence level under a condition of
white noise, while the power peak of PKS 0805-07 is significant at
4.7$\sigma$ (corresponding to the chance probability of
$3\times10^{-4}$\%) under this condition. We also calculated the PSD
slopes for these sources and found that the PSD slope values are
compatible with those are shown in Table \ref{Tab}. Finally, we
computed the chance probability to observe the power peaks in the
GLS periodograms under a condition of power-law noise and found that
4C +01.28 has $0.9$\% chance of being a statistical fluctuation and
PKS 2052-47 has $0.4$\% chance of being a statistical fluctuation.
This strengthens the results obtained from the analysis of
equal-area portions binned with \texttt{HEALPix}.

Other pixels showing periodic behaviours which are not attributed to
a \gray{} source will be discussed. Firstly, the pixel number 10889
satisfies most of the requirements, but does not contain a source
from the 3FGL catalog. The results show that the 656-day periodicity
of the \gray{} signal from pixel 10889 is due to its neighbouring
pixel, 10992, which contains the strong \gray{} source, PKS
2155-304, whose period is about 644 days. Secondly, the pixels with
numbers 5381, 5509, 5510, 5637, 5638, 5765, 5892, 5893, 6021, 6302,
6429, 6430, 6557, 6658, 6684, and 6786, show periodic behaviours
(with a period of about 53 days) caused by the modulation of the
\gray{} signals at the precession period of the orbit of the
\textit{Fermi} spacecraft. Thirdly, the pixels with numbers 5509,
5636, 5766, 6083, 6173, 6430, and 6686, show periodic behaviours
(with a period of about 91 days) caused by the modulation at a
period of 1/4 of a year. This modulation is expected for aperture
photometry for positions in the sky close to the positions of bright
sources. It results from the 4-fold symmetry of the
\textit{Fermi}-LAT PSF and a systematic change in spacecraft
orientation during the course of a
year\footnote{\burl{https://fermi.gsfc.nasa.gov/ssc/data/analysis/LAT\_caveats\_temporal.html}}.
The pixels showing modulated $\gamma$-ray signals due to these two
instrumental effects are shown in Fig. \ref{Fig2} and are
concentrated near the positions of the Vela and Geminga pulsars.
Lastly, no effects caused by the Moon (27.3-day period) and Sun
(1-year period) moving through the sky are detected in this
analysis. This is mostly owing to the choice of long time bins and
small spatial bins used in this analysis. To find more about the
effect due to a \gray{} foreground modulation by the Moon, see
\citet[][]{corbet2013}. To check if the periodic \gray{} foreground
due to the solar \gray{} emission \citep[e.g., for a blazar of 3C
279, see][]{3c279} can affect the conclusion on periodicities of the
seven blazars from Table \ref{Tab}, an additional check is
performed. The closest projected distance from the Sun amongst these
seven sources is 4.6$^{\circ}$ for 4C+01.28 (the 2nd closest is
17$^{\circ}$ for PKS 2155-304). To check the possible influence of
the Sun on \gray{} emission from the pixel containing  4C+01.28, the
time intervals when the solar centre is at a projected distance
closer than 10 degree from this blazar are removed. The subtracted
time intervals include about 5\% of the entire data set. The result
of an analysis of the data set after subtraction is compatible with
the result reported in this paper.

\section{Conclusions}

\textit{Fermi-}LAT capability to observe sources over the entire sky
on a daily basis led to the discovery of orbitally modulated
$\gamma$-ray emission with periods between 2.5 days and 30 days from
several Galactic binaries. Moreover, continuous, ongoing
\textit{Fermi-}LAT all-sky survey started in the summer of 2008
plays a crucial role in studying long-term quasi-periodic
variability of blazars. The discovered quasi-periodic $\gamma$-ray
emission from the blazar, PG 1553+113, with a 2.2 year period
\citep[][]{pg1553} demonstrated the necessity of multi-year
$\gamma$-ray monitoring campaigns of blazars.

In this paper, we performed a systematic search for cyclical sources
of $\gamma$-ray emission. We binned the sky into equal-area pixels
using the \texttt{HEALPix} package and computed the GLS periodograms
in searching for cyclic $\gamma$-ray signals from the sources
belonging to these pixels. Our analysis confirmed the modulated
$\gamma$-ray signals from the three Galactic $\gamma$-ray binaries
and also confirmed the previously claimed quasi-periodicity of
$\gamma$-ray blazars, including PG 1553+113, PKS 2155-304, and BL
Lacertae. The quasi-periodicities of the last two blazars were
claimed by \citet{pks2155p2, bllac} when our manuscript was in
preparation. The advantage of the systematic search for cyclic
sources is that this analysis does not require any pre-selection of
sources. In addition to the previously claimed quasi-periodic
$\gamma$-ray blazars, we found evidence for periodic behaviours of
four other blazars, 4C +01.28, S5 0716+71, PKS 0805-07, and PKS
2052-47. Three of these sources are located at high redshifts and
are potential candidates to binary systems of supermassive black
holes given that the number of such rare systems increases rapidly
with redshift. We conclude that the coming five years of
$\gamma$-ray observations will be decisive in studying the
quasi-periodic variability of these blazars owing to the need for
observations to capture their next several emission cycles.

\section{Acknowledgements}
We are grateful to the referee for the constructive suggestions that
helped us to improve the manuscript. DAP acknowledges support from
the DST/NRF SKA post-graduate bursary initiative. Computations were
performed on the computational facilities belonging to the ALMA Regional
Center Taiwan, Academia Sinica, Taiwan.

\bibliography{refs_v3}

\begin{thebibliography}{50}
\expandafter\ifx\csname natexlab\endcsname\relax\def\natexlab#1{#1}\fi

\bibitem[{{Abdo} {et~al}\mbox{.}(2010){Abdo}, {Ackermann}, {Ajello},
  {Antolini}, {Baldini}, {Ballet}, {Barbiellini}, {Bastieri}, {Bechtol},
  {Bellazzini}, {Berenji}, {Blandford}, {Bloom}, {Bonamente}, {Borgland},
  {Bouvier}, {Bregeon}, {Brez}, {Brigida}, {Bruel}, {Buehler}, {Burnett},
  {Buson}, {Caliandro}, {Cameron}, {Caraveo}, {Carrigan}, {Casandjian},
  {Cavazzuti}, {Cecchi}, {{\c C}elik}, {Chekhtman}, {Cheung}, {Chiang},
  {Ciprini}, {Claus}, {Cohen-Tanugi}, {Cominsky}, {Conrad}, {Costamante},
  {Cutini}, {Dermer}, {de Angelis}, {de Palma}, {Silva}, {Drell}, {Dubois},
  {Dumora}, {Farnier}, {Favuzzi}, {Fegan}, {Focke}, {Fortin}, {Frailis},
  {Fukazawa}, {Funk}, {Fusco}, {Gargano}, {Gasparrini}, {Gehrels}, {Germani},
  {Giebels}, {Giglietto}, {Giommi}, {Giordano}, {Glanzman}, {Godfrey},
  {Grenier}, {Grondin}, {Grove}, {Guiriec}, {Hadasch}, {Hayashida}, {Hays},
  {Healey}, {Horan}, {Hughes}, {Itoh}, {J{\'o}hannesson}, {Johnson}, {Johnson},
  {Kamae}, {Katagiri}, {Kataoka}, {Kawai}, {Kn{\"o}dlseder}, {Kuss}, {Lande},
  {Larsson}, {Latronico}, {Lemoine-Goumard}, {Longo}, {Loparco}, {Lott},
  {Lovellette}, {Lubrano}, {Madejski}, {Makeev}, {Massaro}, {Mazziotta},
  {McEnery}, {Michelson}, {Mitthumsiri}, {Mizuno}, {Moiseev}, {Monte},
  {Monzani}, {Morselli}, {Moskalenko}, {Mueller}, {Murgia}, {Nolan}, {Norris},
  {Nuss}, {Ohno}, {Ohsugi}, {Omodei}, {Orlando}, {Ormes}, {Ozaki}, {Panetta},
  {Parent}, {Pelassa}, {Pepe}, {Pesce-Rollins}, {Piron}, {Porter}, {Rain{\`o}},
  {Rando}, {Razzano}, {Reimer}, {Reimer}, {Ritz}, {Rodriguez}, {Romani},
  {Roth}, {Ryde}, {Sadrozinski}, {Sander}, {Scargle}, {Sgr{\`o}}, {Shaw},
  {Smith}, {Spandre}, {Spinelli}, {Starck}, {Strickman}, {Suson}, {Takahashi},
  {Takahashi}, {Tanaka}, {Thayer}, {Thayer}, {Thompson}, {Tibaldo}, {Torres},
  {Tosti}, {Tramacere}, {Uchiyama}, {Usher}, {Vasileiou}, {Vilchez}, {Vitale},
  {Waite}, {Wallace}, {Wang}, {Winer}, {Wood}, {Yang}, {Ylinen}, \&
  {Ziegler}}]{ferminoise1}
{Abdo} A.~A. {et~al.}, 2010, \apj, 722, 520

\bibitem[{{Abdo} {et~al}\mbox{.}(2009{\natexlab{a}}){Abdo}, {Ackermann},
  {Ajello}, {Atwood}, {Axelsson}, {Baldini}, {Ballet}, {Barbiellini},
  {Bastieri}, {Baughman}, {Bechtol}, {Bellazzini}, {Berenji}, {Blandford},
  {Bloom}, {Bonamente}, {Borgland}, {Bregeon}, {Brez}, {Brigida}, {Bruel},
  {Burnett}, {Caliandro}, {Cameron}, {Caraveo}, {Casandjian}, {Cavazzuti},
  {Cecchi}, {{\c C}elik}, {Charles}, {Chaty}, {Chekhtman}, {Cheung}, {Chiang},
  {Ciprini}, {Claus}, {Cohen-Tanugi}, {Cominsky}, {Conrad}, {Corbel}, {Corbet},
  {Cutini}, {Dermer}, {de Angelis}, {de Luca}, {de Palma}, {Digel}, {Dormody},
  {do Couto e Silva}, {Drell}, {Dubois}, {Dubus}, {Dumora}, {Farnier},
  {Favuzzi}, {Fegan}, {Focke}, {Frailis}, {Fukazawa}, {Funk}, {Fusco},
  {Gargano}, {Gasparrini}, {Gehrels}, {Germani}, {Giebels}, {Giglietto},
  {Giordano}, {Glanzman}, {Godfrey}, {Grenier}, {Grondin}, {Grove},
  {Guillemot}, {Guiriec}, {Hanabata}, {Harding}, {Hayashida}, {Hays}, {Hill},
  {Hughes}, {J{\'o}hannesson}, {Johnson}, {Johnson}, {Johnson}, {Johnson},
  {Kamae}, {Katagiri}, {Kataoka}, {Kawai}, {Kerr}, {Kn{\"o}dlseder}, {Kocian},
  {Kuehn}, {Kuss}, {Lande}, {Larsson}, {Latronico}, {Longo}, {Loparco}, {Lott},
  {Lovellette}, {Lubrano}, {Madejski}, {Makeev}, {Marelli}, {Mazziotta},
  {McEnery}, {Meurer}, {Michelson}, {Mitthumsiri}, {Mizuno}, {Monte},
  {Monzani}, {Morselli}, {Moskalenko}, {Murgia}, {Nolan}, {Nuss}, {Ohsugi},
  {Okumura}, {Omodei}, {Orlando}, {Ormes}, {Paneque}, {Panetta}, {Parent},
  {Pelassa}, {Pepe}, {Pesce-Rollins}, {Piron}, {Porter}, {Rain{\`o}}, {Rando},
  {Ray}, {Razzano}, {Rea}, {Reimer}, {Reimer}, {Reposeur}, {Ritz}, {Rochester},
  {Rodriguez}, {Romani}, {Ryde}, {Sadrozinski}, {Sanchez}, {Sander}, {Saz
  Parkinson}, {Scargle}, {Sgr{\`o}}, {Shaw}, {Sierpowska-Bartosik}, {Siskind},
  {Smith}, {Smith}, {Spandre}, {Spinelli}, {Striani}, {Strickman}, {Suson},
  {Tajima}, {Takahashi}, {Takahashi}, {Tanaka}, {Thayer}, {Thayer}, {Thompson},
  {Tibaldo}, {Torres}, {Tosti}, {Tramacere}, {Uchiyama}, {Usher}, {Vasileiou},
  {Vilchez}, {Vitale}, {Waite}, {Wang}, {Winer}, {Wood}, {Ylinen}, \&
  {Ziegler}}]{bnrLSI61}
{Abdo} A.~A. {et~al.}, 2009{\natexlab{a}}, \apjl, 701, L123

\bibitem[{{Abdo} {et~al}\mbox{.}(2009{\natexlab{b}}){Abdo}, {Ackermann},
  {Ajello}, {Atwood}, {Axelsson}, {Baldini}, {Ballet}, {Barbiellini},
  {Bastieri}, {Baughman}, {Bechtol}, {Bellazzini}, {Berenji}, {Blandford},
  {Bloom}, {Bonamente}, {Borgland}, {Bregeon}, {Brez}, {Brigida}, {Bruel},
  {Burnett}, {Buson}, {Caliandro}, {Cameron}, {Caraveo}, {Casandjian},
  {Cavazzuti}, {Cecchi}, {{\c C}elik}, {Chaty}, {Chekhtman}, {Cheung},
  {Chiang}, {Ciprini}, {Claus}, {Cohen-Tanugi}, {Cominsky}, {Conrad}, {Corbel},
  {Corbet}, {Cutini}, {Dermer}, {de Angelis}, {de Palma}, {Digel}, {Silva},
  {Drell}, {Dubois}, {Dubus}, {Dumora}, {Farnier}, {Favuzzi}, {Fegan}, {Focke},
  {Fortin}, {Frailis}, {Fukazawa}, {Funk}, {Fusco}, {Gargano}, {Gasparrini},
  {Gehrels}, {Germani}, {Giebels}, {Giglietto}, {Giordano}, {Glanzman},
  {Godfrey}, {Grenier}, {Grondin}, {Grove}, {Guillemot}, {Guiriec}, {Hanabata},
  {Harding}, {Hayashida}, {Hays}, {Hill}, {Horan}, {Hughes}, {Jackson},
  {J{\'o}hannesson}, {Johnson}, {Johnson}, {Johnson}, {Kamae}, {Katagiri},
  {Kataoka}, {Kawai}, {Kerr}, {Kn{\"o}dlseder}, {Kocian}, {Kuehn}, {Kuss},
  {Lande}, {Larsson}, {Latronico}, {Lemoine-Goumard}, {Longo}, {Loparco},
  {Lott}, {Lovellette}, {Lubrano}, {Madejski}, {Makeev}, {Marelli},
  {Mazziotta}, {McEnery}, {Meurer}, {Michelson}, {Mitthumsiri}, {Mizuno},
  {Moiseev}, {Monte}, {Monzani}, {Morselli}, {Moskalenko}, {Murgia}, {Nolan},
  {Norris}, {Nuss}, {Ohsugi}, {Omodei}, {Orlando}, {Ormes}, {Ozaki}, {Paneque},
  {Panetta}, {Parent}, {Pelassa}, {Pepe}, {Pesce-Rollins}, {Piron}, {Porter},
  {Rain{\`o}}, {Rando}, {Ray}, {Razzano}, {Rea}, {Reimer}, {Reimer},
  {Reposeur}, {Ritz}, {Rochester}, {Rodriguez}, {Romani}, {Roth}, {Ryde},
  {Sadrozinski}, {Sanchez}, {Sander}, {Saz Parkinson}, {Scargle}, {Sgr{\`o}},
  {Sierpowska-Bartosik}, {Siskind}, {Smith}, {Smith}, {Spandre}, {Spinelli},
  {Strickman}, {Suson}, {Tajima}, {Takahashi}, {Takahashi}, {Tanaka}, {Tanaka},
  {Thayer}, {Thompson}, {Tibaldo}, {Torres}, {Tosti}, {Tramacere}, {Uchiyama},
  {Usher}, {Vasileiou}, {Venter}, {Vilchez}, {Vitale}, {Waite}, {Wallace},
  {Wang}, {Winer}, {Wood}, {Ylinen}, \& {Ziegler}}]{bnrLS5039}
{Abdo} A.~A. {et~al.}, 2009{\natexlab{b}}, \apjl, 706, L56

\bibitem[{{Abdollahi} {et~al}\mbox{.}(2016){Abdollahi}, {Ackermann}, {Ajello},
  {Albert}, {Baldini}, {Ballet}, {Barbiellini}, {Bastieri}, {Becerra Gonzalez},
  {Bellazzini}, {Bissaldi}, {Blandford}, {Bloom}, {Bonino}, {Bottacini},
  {Bregeon}, {Bruel}, {Buehler}, {Buson}, {Cameron}, {Caragiulo}, {Caraveo},
  {Cavazzuti}, {Cecchi}, {Chekhtman}, {Cheung}, {Chiaro}, {Ciprini}, {Conrad},
  {Costantin}, {Costanza}, {Cutini}, {D'Ammando}, {de Palma}, {Desai},
  {Desiante}, {Digel}, {Di Lalla}, {Di Mauro}, {Di Venere}, {Donaggio},
  {Drell}, {Favuzzi}, {Fegan}, {Ferrara}, {Focke}, {Franckowiak}, {Fukazawa},
  {Funk}, {Fusco}, {Gargano}, {Gasparrini}, {Giglietto}, {Giomi}, {Giordano},
  {Giroletti}, {Glanzman}, {Green}, {Grenier}, {Grove}, {Guillemot}, {Guiriec},
  {Hays}, {Horan}, {Jogler}, {J{\'o}hannesson}, {Johnson}, {Kocevski}, {Kuss},
  {La Mura}, {Larsson}, {Latronico}, {Li}, {Longo}, {Loparco}, {Lovellette},
  {Lubrano}, {Magill}, {Maldera}, {Manfreda}, {Mayer}, {Mazziotta},
  {Michelson}, {Mitthumsiri}, {Mizuno}, {Monzani}, {Morselli}, {Moskalenko},
  {Negro}, {Nuss}, {Ohsugi}, {Omodei}, {Orienti}, {Orlando}, {Paliya},
  {Paneque}, {Perkins}, {Persic}, {Pesce-Rollins}, {Petrosian}, {Piron},
  {Porter}, {Principe}, {Rain{\`o}}, {Rando}, {Razzano}, {Razzaque}, {Reimer},
  {Reimer}, {Sgr{\`o}}, {Simone}, {Siskind}, {Spada}, {Spandre}, {Spinelli},
  {Stawarz}, {Suson}, {Takahashi}, {Tanaka}, {Thayer}, {Thompson}, {Torres},
  {Torresi}, {Tosti}, {Troja}, {Vianello}, \& {Wood}}]{2fav}
{Abdollahi} S. {et~al.}, 2016, ArXiv e-prints

\bibitem[{{Acero} {et~al}\mbox{.}(2015){Acero}, {Ackermann}, {Ajello},
  {Albert}, {Atwood}, {Axelsson}, {Baldini}, {Ballet}, {Barbiellini},
  {Bastieri}, {Belfiore}, {Bellazzini}, {Bissaldi}, {Blandford}, {Bloom},
  {Bogart}, {Bonino}, {Bottacini}, {Bregeon}, {Britto}, {Bruel}, {Buehler},
  {Burnett}, {Buson}, {Caliandro}, {Cameron}, {Caputo}, {Caragiulo}, {Caraveo},
  {Casandjian}, {Cavazzuti}, {Charles}, {Chaves}, {Chekhtman}, {Cheung},
  {Chiang}, {Chiaro}, {Ciprini}, {Claus}, {Cohen-Tanugi}, {Cominsky}, {Conrad},
  {Cutini}, {D'Ammando}, {de Angelis}, {DeKlotz}, {de Palma}, {Desiante},
  {Digel}, {Di Venere}, {Drell}, {Dubois}, {Dumora}, {Favuzzi}, {Fegan},
  {Ferrara}, {Finke}, {Franckowiak}, {Fukazawa}, {Funk}, {Fusco}, {Gargano},
  {Gasparrini}, {Giebels}, {Giglietto}, {Giommi}, {Giordano}, {Giroletti},
  {Glanzman}, {Godfrey}, {Grenier}, {Grondin}, {Grove}, {Guillemot}, {Guiriec},
  {Hadasch}, {Harding}, {Hays}, {Hewitt}, {Hill}, {Horan}, {Iafrate}, {Jogler},
  {J{\'o}hannesson}, {Johnson}, {Johnson}, {Johnson}, {Johnson}, {Kamae},
  {Kataoka}, {Katsuta}, {Kuss}, {La Mura}, {Landriu}, {Larsson}, {Latronico},
  {Lemoine-Goumard}, {Li}, {Li}, {Longo}, {Loparco}, {Lott}, {Lovellette},
  {Lubrano}, {Madejski}, {Massaro}, {Mayer}, {Mazziotta}, {McEnery},
  {Michelson}, {Mirabal}, {Mizuno}, {Moiseev}, {Mongelli}, {Monzani},
  {Morselli}, {Moskalenko}, {Murgia}, {Nuss}, {Ohno}, {Ohsugi}, {Omodei},
  {Orienti}, {Orlando}, {Ormes}, {Paneque}, {Panetta}, {Perkins},
  {Pesce-Rollins}, {Piron}, {Pivato}, {Porter}, {Racusin}, {Rando}, {Razzano},
  {Razzaque}, {Reimer}, {Reimer}, {Reposeur}, {Rochester}, {Romani},
  {Salvetti}, {S{\'a}nchez-Conde}, {Saz Parkinson}, {Schulz}, {Siskind},
  {Smith}, {Spada}, {Spandre}, {Spinelli}, {Stephens}, {Strong}, {Suson},
  {Takahashi}, {Takahashi}, {Tanaka}, {Thayer}, {Thayer}, {Thompson},
  {Tibaldo}, {Tibolla}, {Torres}, {Torresi}, {Tosti}, {Troja}, {Van Klaveren},
  {Vianello}, {Winer}, {Wood}, {Wood}, {Zimmer}, \& {Fermi-LAT
  Collaboration}}]{3fgl}
{Acero} F. {et~al.}, 2015, \apjs, 218, 23

\bibitem[{{Ackermann} {et~al}\mbox{.}(2015{\natexlab{a}}){Ackermann}, {Ajello},
  {Albert}, {Atwood}, {Baldini}, {Ballet}, {Barbiellini}, {Bastieri}, {Becerra
  Gonzalez}, {Bellazzini}, {Bissaldi}, {Blandford}, {Bloom}, {Bonino},
  {Bottacini}, {Bregeon}, {Bruel}, {Buehler}, {Buson}, {Caliandro}, {Cameron},
  {Caputo}, {Caragiulo}, {Caraveo}, {Cavazzuti}, {Cecchi}, {Chekhtman},
  {Chiang}, {Chiaro}, {Ciprini}, {Cohen-Tanugi}, {Conrad}, {Cutini},
  {D'Ammando}, {de Angelis}, {de Palma}, {Desiante}, {Di Venere},
  {Dom{\'{\i}}nguez}, {Drell}, {Favuzzi}, {Fegan}, {Ferrara}, {Focke},
  {Fuhrmann}, {Fukazawa}, {Fusco}, {Gargano}, {Gasparrini}, {Giglietto},
  {Giommi}, {Giordano}, {Giroletti}, {Godfrey}, {Green}, {Grenier}, {Grove},
  {Guiriec}, {Harding}, {Hays}, {Hewitt}, {Hill}, {Horan}, {Jogler},
  {J{\'o}hannesson}, {Johnson}, {Kamae}, {Kuss}, {Larsson}, {Latronico}, {Li},
  {Li}, {Longo}, {Loparco}, {Lott}, {Lovellette}, {Lubrano}, {Magill},
  {Maldera}, {Manfreda}, {Max-Moerbeck}, {Mayer}, {Mazziotta}, {McEnery},
  {Michelson}, {Mizuno}, {Monzani}, {Morselli}, {Moskalenko}, {Murgia}, {Nuss},
  {Ohno}, {Ohsugi}, {Ojha}, {Omodei}, {Orlando}, {Ormes}, {Paneque}, {Pearson},
  {Perkins}, {Perri}, {Pesce-Rollins}, {Petrosian}, {Piron}, {Pivato},
  {Porter}, {Rain{\`o}}, {Rando}, {Razzano}, {Readhead}, {Reimer}, {Reimer},
  {Schulz}, {Sgr{\`o}}, {Siskind}, {Spada}, {Spandre}, {Spinelli}, {Suson},
  {Takahashi}, {Thayer}, {Thompson}, {Tibaldo}, {Torres}, {Tosti}, {Troja},
  {Uchiyama}, {Vianello}, {Wood}, {Wood}, {Zimmer}, {Berdyugin}, {Corbet},
  {Hovatta}, {Lindfors}, {Nilsson}, {Reinthal}, {Sillanp{\"a}{\"a}},
  {Stamerra}, {Takalo}, \& {Valtonen}}]{pg1553}
{Ackermann} M. {et~al.}, 2015{\natexlab{a}}, \apjl, 813, L41

\bibitem[{{Ackermann} {et~al}\mbox{.}(2015{\natexlab{b}}){Ackermann}, {Ajello},
  {Albert}, {Atwood}, {Baldini}, {Ballet}, {Barbiellini}, {Bastieri},
  {Bechtol}, {Bellazzini}, {Bissaldi}, {Blandford}, {Bloom}, {Bottacini},
  {Brandt}, {Bregeon}, {Bruel}, {Buehler}, {Buson}, {Caliandro}, {Cameron},
  {Caragiulo}, {Caraveo}, {Cavazzuti}, {Cecchi}, {Charles}, {Chekhtman},
  {Chiang}, {Chiaro}, {Ciprini}, {Claus}, {Cohen-Tanugi}, {Conrad}, {Cuoco},
  {Cutini}, {D'Ammando}, {de Angelis}, {de Palma}, {Dermer}, {Digel}, {Silva},
  {Drell}, {Favuzzi}, {Ferrara}, {Focke}, {Franckowiak}, {Fukazawa}, {Funk},
  {Fusco}, {Gargano}, {Gasparrini}, {Germani}, {Giglietto}, {Giommi},
  {Giordano}, {Giroletti}, {Godfrey}, {Gomez-Vargas}, {Grenier}, {Guiriec},
  {Gustafsson}, {Hadasch}, {Hayashi}, {Hays}, {Hewitt}, {Ippoliti}, {Jogler},
  {J{\'o}hannesson}, {Johnson}, {Johnson}, {Kamae}, {Kataoka},
  {Kn{\"o}dlseder}, {Kuss}, {Larsson}, {Latronico}, {Li}, {Li}, {Longo},
  {Loparco}, {Lott}, {Lovellette}, {Lubrano}, {Madejski}, {Manfreda},
  {Massaro}, {Mayer}, {Mazziotta}, {McEnery}, {Michelson}, {Mitthumsiri},
  {Mizuno}, {Moiseev}, {Monzani}, {Morselli}, {Moskalenko}, {Murgia}, {Nemmen},
  {Nuss}, {Ohsugi}, {Omodei}, {Orlando}, {Ormes}, {Paneque}, {Panetta},
  {Perkins}, {Pesce-Rollins}, {Piron}, {Pivato}, {Porter}, {Rain{\`o}},
  {Rando}, {Razzano}, {Razzaque}, {Reimer}, {Reimer}, {Reposeur}, {Ritz},
  {Romani}, {S{\'a}nchez-Conde}, {Schaal}, {Schulz}, {Sgr{\`o}}, {Siskind},
  {Spandre}, {Spinelli}, {Strong}, {Suson}, {Takahashi}, {Thayer}, {Thayer},
  {Tibaldo}, {Tinivella}, {Torres}, {Tosti}, {Troja}, {Uchiyama}, {Vianello},
  {Werner}, {Winer}, {Wood}, {Wood}, {Zaharijas}, \& {Zimmer}}]{xgal2015}
{Ackermann} M. {et~al.}, 2015{\natexlab{b}}, \apj, 799, 86

\bibitem[{{Ackermann} {et~al}\mbox{.}(2012{\natexlab{a}}){Ackermann}, {Ajello},
  {Atwood}, {Baldini}, {Ballet}, {Barbiellini}, {Bastieri}, {Bechtol},
  {Bellazzini}, {Berenji}, {Blandford}, {Bloom}, {Bonamente}, {Borgland},
  {Brandt}, {Bregeon}, {Brigida}, {Bruel}, {Buehler}, {Buson}, {Caliandro},
  {Cameron}, {Caraveo}, {Cavazzuti}, {Cecchi}, {Charles}, {Chekhtman},
  {Chiang}, {Ciprini}, {Claus}, {Cohen-Tanugi}, {Conrad}, {Cutini}, {de
  Angelis}, {de Palma}, {Dermer}, {Digel}, {Silva}, {Drell}, {Drlica-Wagner},
  {Falletti}, {Favuzzi}, {Fegan}, {Ferrara}, {Focke}, {Fortin}, {Fukazawa},
  {Funk}, {Fusco}, {Gaggero}, {Gargano}, {Germani}, {Giglietto}, {Giordano},
  {Giroletti}, {Glanzman}, {Godfrey}, {Grove}, {Guiriec}, {Gustafsson},
  {Hadasch}, {Hanabata}, {Harding}, {Hayashida}, {Hays}, {Horan}, {Hou},
  {Hughes}, {J{\'o}hannesson}, {Johnson}, {Johnson}, {Kamae}, {Katagiri},
  {Kataoka}, {Kn{\"o}dlseder}, {Kuss}, {Lande}, {Latronico}, {Lee},
  {Lemoine-Goumard}, {Longo}, {Loparco}, {Lott}, {Lovellette}, {Lubrano},
  {Mazziotta}, {McEnery}, {Michelson}, {Mitthumsiri}, {Mizuno}, {Monte},
  {Monzani}, {Morselli}, {Moskalenko}, {Murgia}, {Naumann-Godo}, {Norris},
  {Nuss}, {Ohsugi}, {Okumura}, {Omodei}, {Orlando}, {Ormes}, {Paneque},
  {Panetta}, {Parent}, {Pesce-Rollins}, {Pierbattista}, {Piron}, {Pivato},
  {Porter}, {Rain{\`o}}, {Rando}, {Razzano}, {Razzaque}, {Reimer}, {Reimer},
  {Sadrozinski}, {Sgr{\`o}}, {Siskind}, {Spandre}, {Spinelli}, {Strong},
  {Suson}, {Takahashi}, {Tanaka}, {Thayer}, {Thayer}, {Thompson}, {Tibaldo},
  {Tinivella}, {Torres}, {Tosti}, {Troja}, {Usher}, {Vandenbroucke},
  {Vasileiou}, {Vianello}, {Vitale}, {Waite}, {Wang}, {Winer}, {Wood}, {Wood},
  {Yang}, {Ziegler}, \& {Zimmer}}]{Porter2012}
{Ackermann} M. {et~al.}, 2012{\natexlab{a}}, \apj, 750, 3

\bibitem[{{Ackermann} {et~al}\mbox{.}(2012{\natexlab{b}}){Ackermann}, {Ajello},
  {Ballet}, {Barbiellini}, {Bastieri}, {Belfiore}, {Bellazzini}, {Berenji},
  {Blandford}, {Bloom}, {Bonamente}, {Borgland}, {Bregeon}, {Brigida}, {Bruel},
  {Buehler}, {Buson}, {Caliandro}, {Cameron}, {Caraveo}, {Cavazzuti}, {Cecchi},
  {{\c C}elik}, {Charles}, {Chaty}, {Chekhtman}, {Cheung}, {Chiang}, {Ciprini},
  {}, {Claus}, {Cohen-Tanugi}, {Corbel}, {Corbet}, {Cutini}, {de Luca}, {den
  Hartog}, {de Palma}, {Dermer}, {Digel}, {do Couto e Silva}, {Donato},
  {Drell}, {Drlica-Wagner}, {Dubois}, {Dubus}, {Favuzzi}, {Fegan}, {Ferrara},
  {Focke}, {Fortin}, {Fukazawa}, {Funk}, {Fusco}, {Gargano}, {Gasparrini},
  {Gehrels}, {Germani}, {Giglietto}, {Giordano}, {Giroletti}, {Glanzman},
  {Godfrey}, {Grenier}, {Grove}, {Guiriec}, {Hadasch}, {Hanabata}, {Harding},
  {Hayashida}, {Hays}, {Hill}, {Hughes}, {J{\'o}hannesson}, {Johnson},
  {Johnson}, {Kamae}, {Katagiri}, {Kataoka}, {Kerr}, {Kn{\"o}dlseder}, {Kuss},
  {Lande}, {Longo}, {Loparco}, {Lovellette}, {Lubrano}, {Mazziotta}, {McEnery},
  {Michelson}, {Mitthumsiri}, {Mizuno}, {Monte}, {Monzani}, {Morselli},
  {Moskalenko}, {Murgia}, {Nakamori}, {Naumann-Godo}, {Norris}, {Nuss}, {Ohno},
  {Ohsugi}, {Okumura}, {Omodei}, {Orlando}, {Ozaki}, {Paneque}, {Parent},
  {Pesce-Rollins}, {Pierbattista}, {Piron}, {Pivato}, {Porter}, {Rain{\`o}},
  {Rando}, {Razzano}, {Reimer}, {Reimer}, {Ritz}, {Romani}, {Roth}, {Saz
  Parkinson}, {Sgr{\`o}}, {Siskind}, {Spandre}, {Spinelli}, {Suson},
  {Takahashi}, {Tanaka}, {Thayer}, {Thayer}, {Thompson}, {Tibaldo},
  {Tinivella}, {Torres}, {Tosti}, {Troja}, {Uchiyama}, {Usher},
  {Vandenbroucke}, {Vianello}, {Vitale}, {Waite}, {Winer}, {Wood}, {Wood},
  {Yang}, {Zimmer}, {Coe}, {Di Mille}, {Edwards}, {Filipovi{\'c}}, {Payne},
  {Stevens}, \& {Torres}}]{bnrJ1018}
{Ackermann} M. {et~al.}, 2012{\natexlab{b}}, Science, 335, 189

\bibitem[{{Aharonian} {et~al}\mbox{.}(2006){Aharonian}, {Akhperjanian},
  {Bazer-Bachi}, {Beilicke}, {Benbow}, {Berge}, {Bernl{\"o}hr}, {Boisson},
  {Bolz}, {Borrel}, {Braun}, {Brown}, {B{\"u}hler}, {B{\"u}sching}, {Carrigan},
  {Chadwick}, {Chounet}, {Cornils}, {Costamante}, {Degrange}, {Dickinson},
  {Djannati-Ata{\"i}}, {O'C.~Drury}, {Dubus}, {Egberts}, {Emmanoulopoulos},
  {Espigat}, {Feinstein}, {Ferrero}, {Fiasson}, {Fontaine}, {Funk}, {Funk},
  {F{\"u}{\ss}ling}, {Gallant}, {Giebels}, {Glicenstein}, {Goret},
  {Hadjichristidis}, {Hauser}, {Hauser}, {Heinzelmann}, {Henri}, {Hermann},
  {Hinton}, {Hoffmann}, {Hofmann}, {Holleran}, {Horns}, {Jacholkowska}, {de
  Jager}, {Kendziorra}, {Kh{\'e}lifi}, {Komin}, {Konopelko}, {Kosack},
  {Latham}, {Le Gallou}, {Lemi{\`e}re}, {Lemoine-Goumard}, {Lohse}, {Martin},
  {Martineau-Huynh}, {Marcowith}, {Masterson}, {Maurin}, {McComb}, {Moulin},
  {de Naurois}, {Nedbal}, {Nolan}, {Noutsos}, {Orford}, {Osborne}, {Ouchrif},
  {Panter}, {Pelletier}, {Pita}, {P{\"u}hlhofer}, {Punch}, {Raubenheimer},
  {Raue}, {Rayner}, {Reimer}, {Reimer}, {Ripken}, {Rob}, {Rolland}, {Rowell},
  {Sahakian}, {Santangelo}, {Saug{\'e}}, {Schlenker}, {Schlickeiser},
  {Schr{\"o}der}, {Schwanke}, {Schwarzburg}, {Shalchi}, {Sol}, {Spangler},
  {Spanier}, {Steenkamp}, {Stegmann}, {Superina}, {Tavernet}, {Terrier},
  {Tluczykont}, {van Eldik}, {Vasileiadis}, {Venter}, {Vincent}, {V{\"o}lk},
  {Wagner}, \& {Ward}}]{aharonian2006}
{Aharonian} F. {et~al.}, 2006, \aap, 460, 743

\bibitem[{{Aragona} {et~al}\mbox{.}(2009){Aragona}, {McSwain}, {Grundstrom},
  {Marsh}, {Roettenbacher}, {Hessler}, {Boyajian}, \& {Ray}}]{aragona2009}
{Aragona} C., {McSwain} M.~V., {Grundstrom} E.~D., {Marsh} A.~N.,
  {Roettenbacher} R.~M., {Hessler} K.~M., {Boyajian} T.~S., {Ray} P.~S., 2009,
  \apj, 698, 514

\bibitem[{{Atwood} {et~al}\mbox{.}(2013){Atwood}, {Albert}, {Baldini},
  {Tinivella}, {Bregeon}, {Pesce-Rollins}, {Sgr{\`o}}, {Bruel}, {Charles},
  {Drlica-Wagner}, {Franckowiak}, {Jogler}, {Rochester}, {Usher}, {Wood},
  {Cohen-Tanugi}, \& {S.~Zimmer for the Fermi-LAT Collaboration}}]{atwood2013}
{Atwood} W. {et~al.}, 2013, ArXiv e-prints

\bibitem[{{Atwood} {et~al}\mbox{.}(2009){Atwood}, {Abdo}, {Ackermann},
  {Althouse}, {Anderson}, {Axelsson}, {Baldini}, {Ballet}, {Band},
  {Barbiellini}, \& et~al.}]{atwood2009}
{Atwood} W.~B. {et~al.}, 2009, \apj, 697, 1071

\bibitem[{{Barbiellini} {et~al}\mbox{.}(2014){Barbiellini}, {Bastieri},
  {Bechtol}, {Bellazzini}, {Blandford}, {Borgland}, {Bregeon}, {Bruel},
  {Buehler}, {Buson}, {Caliandro}, {Cameron}, {Caraveo}, {Cavazzuti}, {Cecchi},
  {Chaves}, {Chekhtman}, {Cheung}, {Chiang}, {Ciprini}, {Claus},
  {Cohen-Tanugi}, {D'Ammando}, {de Angelis}, {Dermer}, {Digel}, {Silva},
  {Drell}, {Drlica-Wagner}, {Favuzzi}, {Focke}, {Franckowiak}, {Fukazawa},
  {Fusco}, {Gargano}, {Gasparrini}, {Germani}, {Giglietto}, {Giommi},
  {Giordano}, {Giroletti}, {Glanzman}, {Godfrey}, {Grenier}, {Grove},
  {Guiriec}, {Hadasch}, {Hayashida}, {Hays}, {Hughes}, {Jackson}, {Jogler},
  {Kn{\"o}dlseder}, {Kuss}, {Lande}, {Larsson}, {Longo}, {Loparco},
  {Lovellette}, {Lubrano}, {Mazziotta}, {Mehault}, {Michelson}, {Mizuno},
  {Moiseev}, {Monte}, {Monzani}, {Morselli}, {Moskalenko}, {Murgia}, {Nemmen},
  {Nuss}, {Ohsugi}, {Omodei}, {Orienti}, {Orlando}, {Paneque}, {Perkins},
  {Piron}, {Pivato}, {Prokhorov}, {Rain{\`o}}, {Razzano}, {Razzaque}, {Reimer},
  {Reimer}, {Ritz}, {Romoli}, {S{\'a}nchez-Conde}, {Sanchez}, {Sgr{\`o}},
  {Siskind}, {Spandre}, {Spinelli}, {Takahashi}, {Tanaka}, {Tibaldo},
  {Tinivella}, {Tosti}, {Troja}, {Usher}, {Vandenbroucke}, {Vasileiou},
  {Vianello}, {Vitale}, {Waite}, {Winer}, {Wood}, \& {Yang}}]{3c279}
{Barbiellini} G. {et~al.}, 2014, \apj, 784, 118

\bibitem[{{Begelman} {et~al}\mbox{.}(1980){Begelman}, {Blandford}, \&
  {Rees}}]{Begelman1980}
{Begelman} M.~C., {Blandford} R.~D., {Rees} M.~J., 1980, \nat, 287, 307

\bibitem[{{Bloom} \& {Marscher}(1996)}]{bloom96}
{Bloom} S.~D., {Marscher} A.~P., 1996, \apj, 461, 657

\bibitem[{{Bogdanovi{\'c}}(2015)}]{Bogdanovic2015}
{Bogdanovi{\'c}} T., 2015, in Astrophysics and Space Science Proceedings,
  Vol.~40, Gravitational Wave Astrophysics, {Sopuerta} C.~F., ed., p. 103

\bibitem[{{Callegari} {et~al}\mbox{.}(2009){Callegari}, {Mayer}, {Kazantzidis},
  {Colpi}, {Governato}, {Quinn}, \& {Wadsley}}]{Callegari2009}
{Callegari} S., {Mayer} L., {Kazantzidis} S., {Colpi} M., {Governato} F.,
  {Quinn} T., {Wadsley} J., 2009, \apjl, 696, L89

\bibitem[{{Cavaliere} {et~al}\mbox{.}(2017){Cavaliere}, {Tavani}, \&
  {Vittorini}}]{cavaliere2017}
{Cavaliere} A., {Tavani} M., {Vittorini} V., 2017, \apj, 836, 220

\bibitem[{{Corbet} {et~al}\mbox{.}(2013){Corbet}, {Cheung}, {Kerr}, \&
  {Ray}}]{corbet2013}
{Corbet} R., {Cheung} C.~C., {Kerr} M., {Ray} P.~S., 2013, ArXiv e-prints

\bibitem[{{Corbet} {et~al}\mbox{.}(2016){Corbet}, {Chomiuk}, {Coe}, {Coley},
  {Dubus}, {Edwards}, {Martin}, {McBride}, {Stevens}, {Strader}, {Townsend}, \&
  {Udalski}}]{bnrLMC}
{Corbet} R.~H.~D. {et~al.}, 2016, \apj, 829, 105

\bibitem[{{Dubus}(2013)}]{dubus}
{Dubus} G., 2013, \aapr, 21, 64

\bibitem[{{G{\'o}rski} {et~al}\mbox{.}(2005){G{\'o}rski}, {Hivon}, {Banday},
  {Wandelt}, {Hansen}, {Reinecke}, \& {Bartelmann}}]{Gorsky2005}
{G{\'o}rski} K.~M., {Hivon} E., {Banday} A.~J., {Wandelt} B.~D., {Hansen}
  F.~K., {Reinecke} M., {Bartelmann} M., 2005, \apj, 622, 759

\bibitem[{{Kormendy} \& {Richstone}(1995)}]{Kormendy1995}
{Kormendy} J., {Richstone} D., 1995, \araa, 33, 581

\bibitem[{{Kulkarni} \& {Loeb}(2012)}]{Kulkarni2012}
{Kulkarni} G., {Loeb} A., 2012, \mnras, 422, 1306

\bibitem[{{Liu} {et~al}\mbox{.}(2006){Liu}, {van Paradijs}, \& {van den
  Heuvel}}]{Liu2006}
{Liu} Q.~Z., {van Paradijs} J., {van den Heuvel} E.~P.~J., 2006, \aap, 455,
  1165

\bibitem[{{Liu} {et~al}\mbox{.}(2007){Liu}, {van Paradijs}, \& {van den
  Heuvel}}]{Liu2007}
{Liu} Q.~Z., {van Paradijs} J., {van den Heuvel} E.~P.~J., 2007, \aap, 469, 807

\bibitem[{{Lomb}(1976)}]{Lomb1976}
{Lomb} N.~R., 1976, \apss, 39, 447

\bibitem[{{Maraschi} {et~al}\mbox{.}(1992){Maraschi}, {Ghisellini}, \&
  {Celotti}}]{maraschi92}
{Maraschi} L., {Ghisellini} G., {Celotti} A., 1992, \apjl, 397, L5

\bibitem[{{Mohan} \& {Mangalam}(2015)}]{mohan2015}
{Mohan} P., {Mangalam} A., 2015, \apj, 805, 91

\bibitem[{{Nakagawa} \& {Mori}(2013)}]{ferminoise2}
{Nakagawa} K., {Mori} M., 2013, \apj, 773, 177

\bibitem[{{Prokhorov} \& {Moraghan}(2016)}]{paper1}
{Prokhorov} D.~A., {Moraghan} A., 2016, \mnras, 457, 2433

\bibitem[{{Raiteri} {et~al}\mbox{.}(2015){Raiteri}, {Stamerra}, {Villata},
  {Larionov}, {Acosta-Pulido}, {Ar{\'e}valo}, {Arkharov}, {Bachev},
  {Ben{\'{\i}}tez}, {Bozhilov}, {Borman}, {Buemi}, {Calcidese}, {Carnerero},
  {Carosati}, {Chigladze}, {Damljanovic}, {Di Paola}, {Doroshenko}, {Efimova},
  {Ehgamberdiev}, {Giroletti}, {Gonz{\'a}lez-Morales}, {Grinon-Marin},
  {Grishina}, {Hiriart}, {Ibryamov}, {Klimanov}, {Kopatskaya}, {Kurtanidze},
  {Kurtanidze}, {Kurtenkov}, {Larionova}, {Larionova}, {L{\'a}zaro},
  {L{\"a}hteenm{\"a}ki}, {Leto}, {Markovic}, {Mirzaqulov}, {Mokrushina},
  {Morozova}, {M{\'u}jica}, {Nazarov}, {Nikolashvili}, {Ohlert}, {Ovcharov},
  {Paiano}, {Pastor Yabar}, {Prandini}, {Ramakrishnan}, {Sadun}, {Semkov},
  {Sigua}, {Strigachev}, {Tammi}, {Tornikoski}, {Trigilio}, {Troitskaya},
  {Troitsky}, {Umana}, {Velasco}, \& {Vince}}]{raiteri2015}
{Raiteri} C.~M. {et~al.}, 2015, \mnras, 454, 353

\bibitem[{{Rani} {et~al}\mbox{.}(2013){Rani}, {Krichbaum}, {Fuhrmann},
  {B{\"o}ttcher}, {Lott}, {Aller}, {Aller}, {Angelakis}, {Bach}, {Bastieri},
  {Falcone}, {Fukazawa}, {Gabanyi}, {Gupta}, {Gurwell}, {Itoh}, {Kawabata},
  {Krips}, {L{\"a}hteenm{\"a}ki}, {Liu}, {Marchili}, {Max-Moerbeck},
  {Nestoras}, {Nieppola}, {Quintana-Lacaci}, {Readhead}, {Richards}, {Sasada},
  {Sievers}, {Sokolovsky}, {Stroh}, {Tammi}, {Tornikoski}, {Uemura},
  {Ungerechts}, {Urano}, \& {Zensus}}]{Rani2013}
{Rani} B. {et~al.}, 2013, \aap, 552, A11

\bibitem[{{Rieger}(2004)}]{rieger2004}
{Rieger} F.~M., 2004, \apjl, 615, L5

\bibitem[{{Rieger}(2007)}]{rieger2007}
{Rieger} F.~M., 2007, \apss, 309, 271

\bibitem[{{Sandrinelli} {et~al}\mbox{.}(2014){Sandrinelli}, {Covino}, \&
  {Treves}}]{pks2155p1}
{Sandrinelli} A., {Covino} S., {Treves} A., 2014, \apjl, 793, L1

\bibitem[{{Sandrinelli} {et~al}\mbox{.}(2017){Sandrinelli}, {Covino}, {Treves},
  {Lindfors}, {Raiteri}, {Nilsson}, {Takalo}, {Reinthal}, {Berdyugin}, {Fallah
  Ramazani}, {Kadenius}, {Tuominen}, {Kehusmaa}, {Bachev}, \&
  {Strigachev}}]{bllac}
{Sandrinelli} A. {et~al.}, 2017, ArXiv e-prints

\bibitem[{{Scargle}(1982)}]{Scargle1982}
{Scargle} J.~D., 1982, \apj, 263, 835

\bibitem[{{Sikora} {et~al}\mbox{.}(1994){Sikora}, {Begelman}, \&
  {Rees}}]{sikora1994}
{Sikora} M., {Begelman} M.~C., {Rees} M.~J., 1994, \apj, 421, 153

\bibitem[{{Sobacchi} {et~al}\mbox{.}(2017){Sobacchi}, {Sormani}, \&
  {Stamerra}}]{sobacchi2017}
{Sobacchi} E., {Sormani} M.~C., {Stamerra} A., 2017, \mnras, 465, 161

\bibitem[{{Sobolewska} {et~al}\mbox{.}(2014){Sobolewska}, {Siemiginowska},
  {Kelly}, \& {Nalewajko}}]{ferminoise3}
{Sobolewska} M.~A., {Siemiginowska} A., {Kelly} B.~C., {Nalewajko} K., 2014,
  \apj, 786, 143

\bibitem[{{Tacconi} {et~al}\mbox{.}(2010){Tacconi}, {Genzel}, {Neri}, {Cox},
  {Cooper}, {Shapiro}, {Bolatto}, {Bouch{\'e}}, {Bournaud}, {Burkert},
  {Combes}, {Comerford}, {Davis}, {Schreiber}, {Garcia-Burillo},
  {Gracia-Carpio}, {Lutz}, {Naab}, {Omont}, {Shapley}, {Sternberg}, \&
  {Weiner}}]{Tacconi2010}
{Tacconi} L.~J. {et~al.}, 2010, \nat, 463, 781

\bibitem[{{Tchekhovskoy} {et~al}\mbox{.}(2011){Tchekhovskoy}, {Narayan}, \&
  {McKinney}}]{tchekhovskoy2011}
{Tchekhovskoy} A., {Narayan} R., {McKinney} J.~C., 2011, \mnras, 418, L79

\bibitem[{{Timmer} \& {Koenig}(1995)}]{Timmer1995}
{Timmer} J., {Koenig} M., 1995, \aap, 300, 707

\bibitem[{{VanderPlas} {et~al}\mbox{.}(2012){VanderPlas}, {Connolly},
  {Ivezi{\'c}}, \& {Gray}}]{astroML2012}
{VanderPlas} J.~T., {Connolly} A.~J., {Ivezi{\'c}} {\v Z}., {Gray} A., 2012, in
  Conference on Intelligent Data Understanding (CIDU), pp. 47 --54

\bibitem[{{Volonteri} {et~al}\mbox{.}(2009){Volonteri}, {Miller}, \&
  {Dotti}}]{Volonteri2009}
{Volonteri} M., {Miller} J.~M., {Dotti} M., 2009, \apjl, 703, L86

\bibitem[{{White} \& {Frenk}(1991)}]{White1991}
{White} S.~D.~M., {Frenk} C.~S., 1991, \apj, 379, 52

\bibitem[{{Zechmeister} \& {K{\"u}rster}(2009)}]{Zechmeister2009}
{Zechmeister} M., {K{\"u}rster} M., 2009, \aap, 496, 577

\bibitem[{{Zhang} {et~al}\mbox{.}(2017){Zhang}, {Yan}, {Liao}, \&
  {Wang}}]{pks2155p2}
{Zhang} P.-f., {Yan} D.-h., {Liao} N.-h., {Wang} J.-c., 2017, \apj, 835, 260

\end{thebibliography}

\end{document}